\documentclass[a4paper,fleqn,usenatbib]{mnras}

\usepackage{newtxtext,newtxmath}
 
\usepackage[T1]{fontenc}
\usepackage{ae,aecompl}
\usepackage{textcomp}

\usepackage{graphicx}	
\usepackage{amsmath}	
\usepackage{amssymb}	
\usepackage{pdfpages}   
\usepackage{acro}       
\usepackage{cleveref}

\DeclareAcronym{gr}{
  short = GR ,
  long = General Relativity ,
  short-plural = ,
}

\DeclareAcronym{smbh}{
  short = SMBH ,
  long = Supermassive Black Hole ,
  short-plural = s,
}
\DeclareAcronym{gw}{
  short = GW ,
  long = Gravitational Wave ,
  short-plural = ,
}

\DeclareAcronym{lisa}{
  short = {\it LISA} ,
  long = {\it Laser Interferometric  Space  Antenna},
  short-plural = ,
}

\DeclareAcronym{pn}{
  short = PN ,
  long = post-Newtonian ,
  short-plural = ,
}
\DeclareAcronym{bh}{
  short = BH ,
  long = Black Hole ,
  short-plural = s,
  long-plural = s,
}

\DeclareAcronym{ligo}{
  short = LIGO ,
  long = Laser Interferometer Gravitational-Wave Observatory ,
  short-plural = ,
}
\DeclareAcronym{msp}{
  short = MSP ,
  long = millisecond pulsar ,
  short-plural = s,
  long-plural = s,
}
\DeclareAcronym{emrb}{
  short = EMRB ,
  long = extreme-mass-ratio binary ,
  short-plural = s,
  long-plural-form = extreme-mass-ratio binaries,
}
\DeclareAcronym{emri}{
  short = EMRI ,
  long = extreme-mass-ratio-inspiral ,
  short-plural = s,
}
\DeclareAcronym{mpd}{
  short = MPD ,
  long = Mathisson-Papapetrou-Dixon ,
  short-plural = ,  
}
\DeclareAcronym{eom}{
  short = EOMs ,
  long = equations of motion ,
}
\DeclareAcronym{gc}{
  short = GC ,
  long = globular cluster ,
  short-plural = s,
  long-plural = s,
}

\newcommand{\kms}{\,{\rm km}/{\rm s}}
\newcommand{\solarmass}{\,{\rm M}_\odot}

\title[Scattering of a neutron star by a black hole]
{Relativistic scattering of a fast spinning neutron star   
  by a massive black hole}

\author[K. J. Li et al.]{
Kaye Jiale Li$^{1,2}$\thanks{E-mail: j-li.19@ucl.ac.uk (KJL), kinwah.wu@ucl.ac.uk (KW), pkleung@phy.cuhk.edu.hk (PKL), dinesh.singh@uregina.ca (DS)},
Kinwah Wu$^{1}$,
Po Kin Leung$^{2}$
and Dinesh Singh$^{3}$
\\
$^{1}$Mullard Space Science Laboratory, University College London, Holmbury St Mary, Surrey, RH5 6NT, United Kingdom \\
$^{2}$Department of Physics, Chinese University of Hong Kong, Shatin, NT, Hong Kong SAR, China \\ 
$^{3}$Department of Physics, University of Regina, Regina, SK S4S 0A2, Canada
}

\date{Accepted 2021 October 5. Received 2021 October 4; in original form 2021 February 3}

\pubyear{2021}

\begin{document}
\label{firstpage}
\pagerange{\pageref{firstpage}--\pageref{lastpage}}
\maketitle

\begin{abstract}
The orbital dynamics 
  of fast spinning neutron stars encountering a massive \ac{bh} with unbounded orbits are investigated 
  using the quadratic-in-spin \ac{mpd} formulation. 
We consider the motion of the spinning neutron stars
  with astrophysically relevant speed in the gravity field of the BH. 
For such slow-speed scattering, the hyperbolic orbits followed by these neutron stars all have
  near the $e=1$ eccentricity, and have distinct properties compared with those of  $e \gg 1$.  
We have found that 
  compared with geodesic motion, 
  the spin-orbit and spin-spin coupling will lead to a variation of scattering angles 
  at spatial infinity,
  and this variation is more prominent for slow-speed scattering than fast-speed scattering. 
Such a variation leads to an observable difference in pulse-arrival-time within a few hours of observation, 
  and up to a few days or months for larger BH masses or longer spinning periods.  
Such a relativistic pulsar-BH system also emits a burst of gravitational waves (GWs) in the sensitivity band of LISA, 
  and for optimal settings, can be seen up to $100\,{\rm Mpc}$ away. 
A radio follow up of such a GW burst with SKA or FAST 
  will allow for measuring the orbital parameters with high accuracy and 
  testing the predictions of \ac{gr}. 
\end{abstract}




\section{Introduction}
\label{sec:intro} 

Pulsars are either fast spinning young neutron stars with a strong magnetic field 
  ($\sim 10^{12} - 10^{13}\;\!{\rm G}$) 
  or old recycled neutron stars with a weaker magnetic field ($\sim 10^{9}\;\!{\rm G}$) 
  \citep[see e.g.][]{Lorimer2008}.   
The latter are \acp{msp}, 
  whose spin periods range from $\sim 1 - 10\;\!{\rm ms}$ 
  and are extremely stable,   
  with a drift of roughly 1 pulse period on a Hubble timescale  
  \citep[see e.g.][]{Manchester2017}. 
\acp{msp} are mass gyros, 
  and the stability of their spin periods   
  make them very high-precision timing instruments.  
Relativistic binaries containing an \ac{msp} orbiting 
  around a \ac{bh}, 
  i.e.\, \ac{msp} - \ac{bh} binaries, 
  are particularly useful for the investigation of fundamental physics, 
  where accurate, reliable measurements are essential.
The two subclasses of the \ac{msp} - \ac{bh} binaries, 
  the \ac{emrb} 
  and the \ac{emri} systems,   
  each contain a massive \ac{bh}.    
These systems are natural GW sources,  
  expected to be detected 
  by the \ac{lisa} 
  \citep[see][]{Amaro-Seoane2007,Gair2017}. 
As \acp{msp} are  radio sources, 
  an ``observable'' electromagnetic counterpart 
  will be present  
  if an \ac{emrb}/\ac{emri} GW event occurs,  
  allowing 
  high-precision measurements to be made 
  in the electromagnetic messenger domain 
  and in the gravitational messenger domain 
  independently. 
In the theoretical perspective,  
  the fact that neutron stars 
  have a relatively narrow mass range   
  \citep[see][]{Lattimer2012,Ozel2016} 
  reduces one important system parameter  
  in the orbital and spin dynamical modelling.   
Moreover, as the mass of a neutron star 
  is small in comparison with the massive nuclear \ac{bh},  
  the \ac{msp} can be treated as a test object. 

In the test-object limit, 
  the motion of a spinning secondary in the gravity field 
  of a massive black hole is governed by 
  the MPD equations.
Most of the current studies \citep[e.g.][]{Semerak1999,Yunes2011,Singh2014,Kimpson2020a,Kimpson2020b} 
  using MPD equations have put focus on bounded systems and their orbital dynamics.
Also a circular or a quasi-circular orbit approximation 
  is often adopted 
  \citep[e.g.][]{Bini2005,Han2010,Velandia2018,Chen2019},     
  where the spin-spin and spin-orbit couplings 
  are treated as independent separated components 
  \citep[see also the application of same deposition 
   in the study of binary neutron stars around a black hole 
   in][]{Remmen2013}. 
Compared to the bounded orbits, 
  hyperbolic orbits admit the advantage of having clean 
  non-degenerate observables. 
Among them the most well recognised is the scattering angle $\chi$ ($\equiv \Delta \phi$ in this paper).
The correction to the scattering angle due to the 
  spin of the secondary has been widely studied.
For example, \citet*{Bini2017a,Bini2017b} calculated the analytical formula
  (to the first order in spin) of the orbit of a spinning particle
  on the equatorial plane around a massive \ac{bh} using the \ac{mpd} equations.
\citet{Bini2018} studied the effects of spin and spin-induced quadrupolar moments on the scattering angle
  in the high-energy limit.
MPD equations are also adopted in the theoretical modelling of 
  scattering system by \citet*{Vines2016,Vines2018,Vines2019,Antonelli2020}, etc. 
While analytical analysis of the dynamics of spinning binary 
  tend to restrict the orbits to equatorial plane, 
  where the spin is aligned or anti-aligned with the orbital angular momentum, 
  the orbits can be complicated and even chaotic \citep[see e.g.][]{Suzuki1997,Hartl2003a,Hartl2003b,Kao2005,Zelenka2019,Witzany2019a,Zelenka2020} 
  when the spin of the secondary is not aligned with the orbital angular momentum.
The rich dynamical features of non-equatorial motions are 
  studied by \citet*{Singh2014,Han2017,Witzany2019b,Li2019,Kimpson2019a,Keresztes2019,Keresztes2020}, etc.
There are also a few studies \citep[e.g.][]{Hansen1972,Majar2010,DeVittori2014}
  focusing on the gravitational radiation from unbounded systems.

Here we investigate in the astrophysical context 
  the dynamics of the unbounded \ac{emrb}, 
  where an \ac{msp} (i.e., spinning mass gyro with a radio beaming beacon)  
  interacts with a stationary space-time provided by a massive \ac{bh}. 
The focus of the work is 
  on the non-linearity and complexity arisen from the spin-couplings 
  in a fly-by encounter, i.e. 
  the \ac{msp} is in an unbounded orbit with respect to the \ac{bh}.  
The \ac{mpd} formulation is adopted 
 to derive the orbital and spin evolutionary equations  
  of the \ac{msp}. 
By solving the orbital and spin evolutionary equation 
  we determine the multi-messenger signatures of these system. 
We organise the paper as follows. 
In \S2 we present the \ac{mpd} \ac{eom} for
  the astrophysically relevant parameter space that concerns this paper.  
In \S3 we extend the numerical solution to radial infinity to compare the scattering angles
  with the linear-in-spin analytical formula from \citet{Bini2017a} and \citet{Bini2017b},
  show the complex orbital dynamics of the scattering, 
  and investigate the detectability of the spin's effects in such an \ac{emrb} system;
In \S4 we comment on the existence of such \ac{emrb} systems and 
  the implications in astrophysics and multi-messenger observations, after which we conclude the paper in \S5.  

\section{unbounded scattering between neutron stars and a massive BH} 
\label{sec:hyperbolic}

We adopt a $[\, - \;\! +\;\! +\;\! +\, ]$  metric signature 
  and a natural unit system 
  with unity speed of light $c$ and gravitational constant $G$ 
  (i.e.\, $c= G =1$). 
The Schwarzschild radius of a \ac{bh} is therefore $r_{\rm s} = 2M$. 
The pulsar is a spinning test object, 
  and its mass is fixed to be $1.5~{\rm M}_\odot$, 
  a value typical for a neutron star. 
The dimensionless spin of the neutron star with period $P_{\rm ns}$ is set to be 
\begin{equation}
   \hat{s} =  \frac{s}{m M} = \left( \frac{1 {\rm ms}}{P_{\rm ns}}  \right)
   \left( \frac{10^3~{\rm M}_\odot }{M}  \right)
   \times 5.68 \times 10^{-4} \ , 
\end{equation}
under the assumption that the pulsar is a solid sphere with uniform density and radius $10\,{\rm km}$.
We refer to $P_{\rm ns}=1\, {\rm ms}$ pulsars as \acp{msp}.  
  
We adopt the \ac{mpd} formulation up to quadrupole interaction. 
The orbital and spin evolutions of the pulsar are governed by the equations:
\begin{equation}  
  \dot{p}^{\mu}
    =  -\frac{1}{2}\;\! {R^{\mu}}_{\nu\alpha \beta}u^\nu s^{\alpha \beta} 
  + {\cal F}^{\mu}  \ ; 
\label{eq-MPD_a1}
\end{equation} 
\vspace*{-12pt}
\begin{equation}  
\dot{s}^{\mu \nu}
  =  p^\mu u^\nu - p^\nu u^\mu   
 + {\cal T}^{\mu \nu} 
\label{eq-MPD_a2}
\end{equation} 
\vspace*{-12pt}
\begin{equation}
\begin{aligned}
 \mathcal{F}^{\mu}  &  \equiv - \frac{1}{6} J^{\alpha \beta \gamma \sigma} \nabla^{\mu} R_{\alpha \beta \gamma \sigma}  \ , \\
 \mathcal{T}^{\mu \nu} & \equiv \frac{4}{3} J^{\alpha \beta \gamma[\mu} R^{\nu]} {}_{\gamma \alpha \beta}  \ .
\end{aligned}
\end{equation}
\citep[see][]{Mathisson1937,Papapetrou1951,Dixon1964},
  where $u^\mu = {\rm d}x^\mu/{\rm d}\tau$
  is the unit tangent vector along the worldline of the pulsar's centre of mass.
The over dot represents the covariant derivative along this worldline, 
  i.e. $\dot{p}^{\mu} \equiv u^{\nu} \nabla_{\nu} p^{\mu}$. 
The Dixon force and torque are $\mathcal{F}^{\mu}$ 
and $\mathcal{T}^{\mu \nu} $, respectively, and  
$J^{\alpha \beta \gamma \sigma}$ is the quadrupole tensor of the neutron star.
We adopt the spin-induced quadrupole tensor as used in \citet{Steinhoff2011}.
\begin{equation}
\begin{aligned}
 J^{\alpha \beta \gamma \sigma}  & = 
   4 \upsilon^{[\alpha} \chi (\upsilon)^{\beta][\gamma} \upsilon^{\sigma]}  \ , {\rm with}  \\
  \chi(\upsilon)^{\beta \gamma} & = \frac{3}{4} \frac{C_Q}{m} 
 \left[ s^{\beta} s^{\gamma} - \frac{1}{3} s^2 \left( g^{\beta \gamma} + \upsilon^{\beta} \upsilon^{\gamma} \right) \right]  \ .
\end{aligned}
\end{equation}
and $\upsilon^{\alpha} \equiv p^{\alpha}/m$,
where $m \equiv \sqrt{-p^{\mu}p_{\mu}}$.
$C_Q$ is the polarizability constant and its value depends on the equation of state of the object. 
It is normalised such that $C_Q=1$ corresponds to a BH.
For neutron star, the value of $C_Q$ is chosen to be between $3.1$ and $7.4$ 
\citep{Laarakkers1999}\footnote{Notice that they did not study $1.5 \solarmass$ but 
  $1.4 \solarmass$ and $1.6 \solarmass$ neutron stars.}.
\citet{Urbanec2013} 
found a value between $5-6$. 
We take $C_Q$ to be $6$ and argue that the main findings in this work are dominated by spin-couplings.
The covariant derivatives 
  are taken with respect to a background metric,
  which is provided by a Kerr \ac{bh}, 
  with the line element given by 
\begin{equation} 
\begin{aligned} 
  - {\rm d}\tau^2 = & - \left(1- \frac{2Mr}{\Sigma}\right) {\rm d} t^2 -  
     \frac{4a M r \sin^2 \theta}{\Sigma}\;\! {\rm d} t \;\!{\rm d} \phi   
      \\  
    &  + \frac{\Sigma}{\Delta}\;\! {\rm d} r^2 + \Sigma\;\! {\rm d}\theta^2     
    +\left(r^2+a^2 +\frac{2a^2Mr \sin^2\theta}{\Sigma} \right)   \\
   & \times      \sin^2\theta \;\! {\rm d} \phi^2  \   ,  
\end{aligned} 
\label{eq-Kerr}  
\end{equation} 
   in Boyer-Lindquist coordinates.  
Here $\Sigma = r^2 + a^2\cos^2 \theta$, 
   $\Delta = r^2 - 2Mr +a^2$ and 
   $(r,\theta, \phi)$ represent the spatial 3-vector 
   in (pseudo-)spherical polar coordinates with the black-hole centre as the origin. 

Here we consider 
  the Tulczyjew-Dixon (TD) spin supplementary condition  
   \citep[see][]{Tulczyjew1959,Deriglazov2017}, i.e.   
\begin{equation} 
  s^{\mu \nu} p_{\nu} = 0   \ , 
\end{equation}  
and use the spin-vector to simplify the \ac{eom}:
\begin{equation}  
     s_\mu = - \frac{1}{2m}\;\! \epsilon_{\mu \nu \alpha \beta}p^\nu s^{\alpha\beta} \ ; 
\label{eq-spin_1}
\end{equation} 
\vspace*{-12pt}
\begin{equation} 
   s^{\mu \nu } = \frac{1}{ m}\;\! \epsilon^{\mu \nu \alpha \beta} p_\alpha s_\beta  \ ,  
\label{eq-spin_2}  
\end{equation}  
  with Levi-Civita tensor 
   $\epsilon_{\mu \nu \alpha \beta} = \sqrt{-\;\! g\;\!}\;\! \sigma_{\mu \nu \alpha \beta}$  
   adopting the $\sigma_{0123} = +1$ permutation. 

Under this spin supplementary condition, the \ac{eom} now
  become
\begin{equation}  
\begin{aligned} \label{eq:mpd-momentum}
\frac{{\rm d} p^\alpha}{{\rm d}\tau} = & - {\Gamma^\alpha}_{\mu \nu} p^\mu u^\nu 
   + \lambda \left(   -\frac{1}{2}\;\! {R^{\alpha}}_{\beta\mu \nu}u^{\beta} s^{\mu \nu} 
  + {\cal F}^{\alpha}    \right) \ ; \\ 
\end{aligned} 
\end{equation}    
\vspace*{-12pt}
\begin{equation}  
\begin{aligned} \label{eq:mpd-spin}
\frac{{\rm d} s^\alpha}{{\rm d}\tau} = & - {\Gamma^\alpha}_{\mu \nu} s^\mu u^\nu  
 +  \lambda  \bigg[ \left(  -\frac{1}{2 m^2}\;\! R_{\gamma\beta\mu \nu}u^{\beta} s^{\mu \nu} 
 + {\cal F}_{\gamma}  \right) s^{\gamma} p^{\alpha}  \\ 
 &  - \frac{1}{2m} {\epsilon^{\alpha}}_{\beta \mu \nu} p^{\beta} {\cal T}^{\mu \nu}  
\bigg]  \ ; \\ 
\end{aligned} 
\end{equation}    
where we follow the method in \citet{Singh2005} and \citet{Singh2014} to introduce the
  parameter switch $\lambda$ 
  into the \ac{mpd} equations. 
The spin-curvature coupling is included when  
   setting $\lambda =1$.
Further, quadrupole-curvature coupling is included when both $C_Q$ and $\lambda$ are non-zero. 
With these two switches,
  we can compare the difference in the dynamics 
  between the presence and the absence of spin-curvature and quadrupole-curvature couplings.

It remains to determine the 4-velocity.
We follow the procedure of \citet*{Semerak1999,Han2017} to first contract Eq.~\ref{eq-MPD_a2} with $p_{\nu}$,
  making use of the relation 
$R_{\nu \sigma \alpha \beta} s^{\mu \nu} s^{\sigma \beta} = - R_{\nu \sigma \alpha \beta} s^{\nu \sigma} s^{\mu \beta}/2$,
  and then making use of the TD spin supplementary condition to replace $p_{\mu} \dot{s}^{\nu \mu}$
  with $\dot{p}_{\mu} s^{\mu \nu}$. 
Finally, the 4-velocity takes the following form:
\begin{equation}  
\begin{aligned} \label{eq:mpd-4velocity}
\frac{{\rm d} x^\alpha}{{\rm d}\tau} = & u^{\alpha} = \frac{m_u}{m^2} \left( p^{\alpha} - \dot{p}_{\nu} s^{\alpha \nu}\right) - \frac{1}{m^2} p_{\nu} F^{\alpha \nu} \ , \\
 \dot{p}_{\mu} s^{\eta \mu} = & \frac{1}{4 m^{2}+R_{\mu \nu \alpha \beta} s^{\alpha \beta} s^{\mu v}} 
\bigg(
-2 m_u R_{\mu \nu \alpha \beta}  s^{\alpha \beta} p^{v} s^{\eta \mu} \\
 & +2 R_{\mu \nu \alpha \beta } s^{\alpha \beta} p_{\delta} {\mathcal T}^{\nu \delta} s^{\eta \mu}-4 m^{2} {\mathcal F}_{\mu} s^{\eta \mu}
\bigg)
\ , \\
\end{aligned} 
\end{equation}    
where $m_u \equiv -p^{\mu} u_{\mu}$ is the mass measured by an observer with 4-velocity $u^{\mu}$.
These two masses $m$ and $m_u$ are, in general, not the same.  
As a set of non-linear equations, (\ref{eq:mpd-4velocity})
is underdetermined, but a unique solution can be obtained by use of the normalisation condition $u_{\mu} u^{\mu} = -1$.
The MPD formula does not require $u^{\mu}$ to be unit-normal.
For example, \citet{Ehlers1977} assumed $u_{\mu} p^{\mu} = -m$ (they used $\dot{z}^{\mu}$ instead of $u_{\mu}$)
  by scaling the time parameter of the curve.
Therefore, requiring $u^{\mu}$ to be either unit-normal or to satisfy $u_{\mu} p^{\mu} = -m$ are 
  two distinct choices.
In this study we have chosen $u^{\mu}$ to be unit-normal, such that the time parameter becomes the
  proper time.

There are several constants of motion related to the MPD equation under TD spin supplementary condition.
The Killing vector field $\xi^{\mu} = (1, 0,0,0)$ and $\eta^{\mu} = (0,0,0,1)$ give rise to
  the conservation of energy and angular momentum: 
\begin{equation}  
\begin{aligned} 
E & =-\xi_{\alpha} p^{\alpha}+\frac{1}{2} s^{\alpha \beta} \nabla_{\beta} \xi_{\alpha} \ , \\
J & =\eta_{\alpha} p^{\alpha}-\frac{1}{2} s^{\alpha \beta} \nabla_{\beta} \eta_{\alpha} \ , 
\end{aligned} 
\end{equation}    
where $ \nabla_{\beta} \xi_{\alpha} = g_{t [\alpha, \beta]}$ and $\nabla_{\beta} \eta_{\alpha} = g_{\phi [\alpha, \beta]}$.
There is a further constant of motion that is conserved at first order in spin given by the Killing-Yano tensor \citep{Rudiger1981,Rudiger1983,Witzany2019b}:
\begin{equation}  
\begin{aligned} 
K & = Y_{\mu \chi} {Y_\nu}^{\chi} p^{\mu} p^{\nu} - 2\lambda p^{\mu} s^{\rho \sigma}
      \left( Y_{\mu \rho; \kappa} {Y^{\kappa}}_{\sigma} + Y_{\rho \sigma;\kappa} {Y^{\kappa}_{\mu}} \right) \ ,  
\end{aligned} 
\end{equation}    
where $Y_{\mu \nu}$ is the Killing-Yano tensor with components listed in Eq.~2 of \citet{Witzany2019b}.
The dynamical mass $m$ is a constant of motion at dipole order, 
  but varies as $\mathcal{O}(s^2)$ at the quadrupole order. 
A mass-like definition 
\begin{equation}  
\begin{aligned} 
m_J \equiv m - \frac{1}{6} R_{\alpha \beta \mu \nu} J^{\alpha \beta \mu \nu} \ , \\
\end{aligned} 
\end{equation}    
\citep{Steinhoff2011} varies as $\mathcal{O}(s^4)$ at the quadrupole order. 
Although the magnitude of spin
\begin{equation}  
\begin{aligned} 
s^2 \equiv \frac{1}{2} s^{\mu \nu} s_{\mu \nu} \equiv s^{\mu} s_{\mu} \\
\end{aligned} 
\end{equation}    
is not conserved for a general quadrupole tensor, it is conserved for this ansatz spin-induced quadrupole tensor.

The initial conditions are set to be $x^\mu = (0, 10^4 \, M, \pi/2, 0)$,
  corresponding to a pulsar on $x$-axis, at $t=\tau=0$.  
At such large radial distance, we have $p^\mu = m u^\mu$  with $m = 1.5 \solarmass$,
  and therefore $p_t \equiv - m E_0$ and $p_\phi \equiv m J_0$, 
  where $E_0$ and $J_0$ are defined in Eq.~\ref{eq:geodesicsELz}.    
Further, the test object is initially moving on the equatorial plane
  and we have $p^\theta = 0$. 
Then, $p^r$ can be determined by the condition $p^\mu p_\mu = - m^2$.  
The initial values of the 4-spin are determined by $\theta_{\rm S}$ and $\phi_{\rm S}$ 
  which are defined with repect to a Cartesian coordinate in the \ac{msp}'s local tetrad frame,
  which is approximately equivalent to the global Cartesian coordinate at large radial distance.  
For example, when $\theta_{\rm S}=90^\circ, \, \phi_{\rm S}=0^\circ$, the 3-spin is in radial direction,
  and when $\theta_{\rm S}=0^\circ$, the 3-spin is parallel to the angular momentum.

\subsection{Hyperbolic orbits of MSPs around \ac{bh}} 
\label{subsec:eccentricity}  

For a massive nuclear \ac{bh}, stellar interactions dominate outside the influence radius of the \ac{bh}: 
\begin{equation}
\begin{aligned}
r_{\rm inf} \equiv \frac{G M}{\sigma^2} \ ,
\end{aligned}
\end{equation}
where $\sigma$ is the velocity dispersion. 
If intermediate-mass \acp{bh} do follow the M-$\sigma$ relation, 
  then $\sigma \approx 18 \kms $ for an $M=10^3 \, \solarmass$ \ac{bh} \citep{McConnell2011}, 
  giving rise to an influence radius of about $ 0.01 {\rm pc}$,
  while $M=10^5 \solarmass$ \acp{bh} have an influence radius of about $0.2 \, {\rm pc}$. 
Measurements of \acp{gc} give similar results, ranging from $3 \kms $ to $20 \kms$ \citep{Baumgardt2018}. 

If a star is scattered into a hyperbolic orbit around the \ac{bh}, 
  the star's orbit and periapsis distance are uniquely described by its initial velocity (velocity at $r\to \infty$, we denote $v_\infty$) 
  and impact parameter $b$ (or angular momentum $L_z$). 
Stars with smaller initial velocities can get closer to the \ac{bh} even when the impact parameter is large, 
  because of their small inertia. 
The impact parameter is related to the angular momentum $J_0$ and $v_\infty$ via: 
\begin{equation}
\begin{aligned} \label{eq:impactparameter}
b \equiv & \frac{J_0}{E_0 v_\infty} \approx \frac{\sqrt{ 2 M r_{\min}} }{v_\infty } \ , \\
\approx & 4\, {\rm au} \times  
\bigg( \frac{r_{\min}}{100\, M} \bigg)^{1/2} 
\bigg( \frac{M       }{10^3 \solarmass} \bigg)
\bigg( \frac{10 \kms}{v_\infty} \bigg) 
\end{aligned}
\end{equation}
For a $M=10^3 \, \solarmass$ \ac{bh},
  the impact parameter is in general greater than $1\,{\rm au}$ for a close fly-by with $r_{\min} \le 100\, M$ 
  when $v_\infty \le 10 \kms$. 
As the scattering cross section is proportional to the square of $b$, 
  lower velocities and a larger \ac{bh}'s mass will lead to larger scattering cross sections.

While young pulsars are high-speed stars (a mean velocity of $\sim 450 \kms $) 
  that could probably escape the galactic potential \citep{Lyne1994},
  old pulsars (some of which are \ac{msp}s) that have undergone 
  accretion processes are believed to 
  move with much smaller speed, and reach equilibrium during their interactions with the environments.
The mean velocity of \ac{msp} is generally believed to be around 
$100\, \kms$ \citep{Lyne1998,Hobbs2005,Malov2007,Gonzalez2011} 
  while some expect lower values \citep{Cordes1997,Hooper2013}. 

In this work, we consider a typical speed 
   $v_\infty = 5-10 \kms$ (\acp{gc}), 
  $50 \kms$ (small galaxies) and $100  \kms $ (large galaxies) for the \ac{msp}. 
We also consider unrealistic cases with $v_\infty=0.1 c $ and $c/3$ for comparison. 
The central intermediate mass \ac{bh} is set to be of mass $M = 10^3 \, \solarmass$, 
  but the results are applicable for more massive \acp{bh} as well (with an approximate dependency on the mass ratio $m/M$).  
We mainly consider a close encounter of $10 \, M < r_{\rm min} < 200 \, M$, 
  but extend the study to a larger parameter space to explore the full extent of observational possibilities.

We notice that, when comparing the results of geodesic motion with spin-coupled motions, 
  there are subtle differences in comparing orbits with the same 
  $E_0, J_0$ and orbits with the same $p$ (semi-latus rectum), $e$ 
  as pointed out by \citet{Bini2017b}. 
Here, we choose to compare orbits with the same $E_0, J_0$ defined at past infinity, 
  because under the influence of spin, 
  these two values are no longer 
  conserved due to the break down of symmetry. 
The reason is that, 
  if an \ac{msp} is found on a hyperbolic orbit approaching 
  the \ac{bh} in the weak field, 
  the motion under the geodesic equation and 
  the motion under the \ac{mpd} equations are indistinguishable if the \ac{msp} has the same $E_0$ and $J_0$ at past infinity.
But orbits with the same $p,e$ under different \ac{eom} are distinguishable even in the weak field,
  and therefore these orbits will not be considered to be the same.
Hence the difference in the predicted orbits under different \ac{eom} lack sufficient physical meaning.

\section{Spin-orbit coupling and spin-curvature coupling}
\label{sec:coupling}

\subsection{Relativistic hyperbolic orbit}

\begin{figure}
    \centering
    \vspace*{-0cm}
    \includegraphics[width=1\columnwidth]{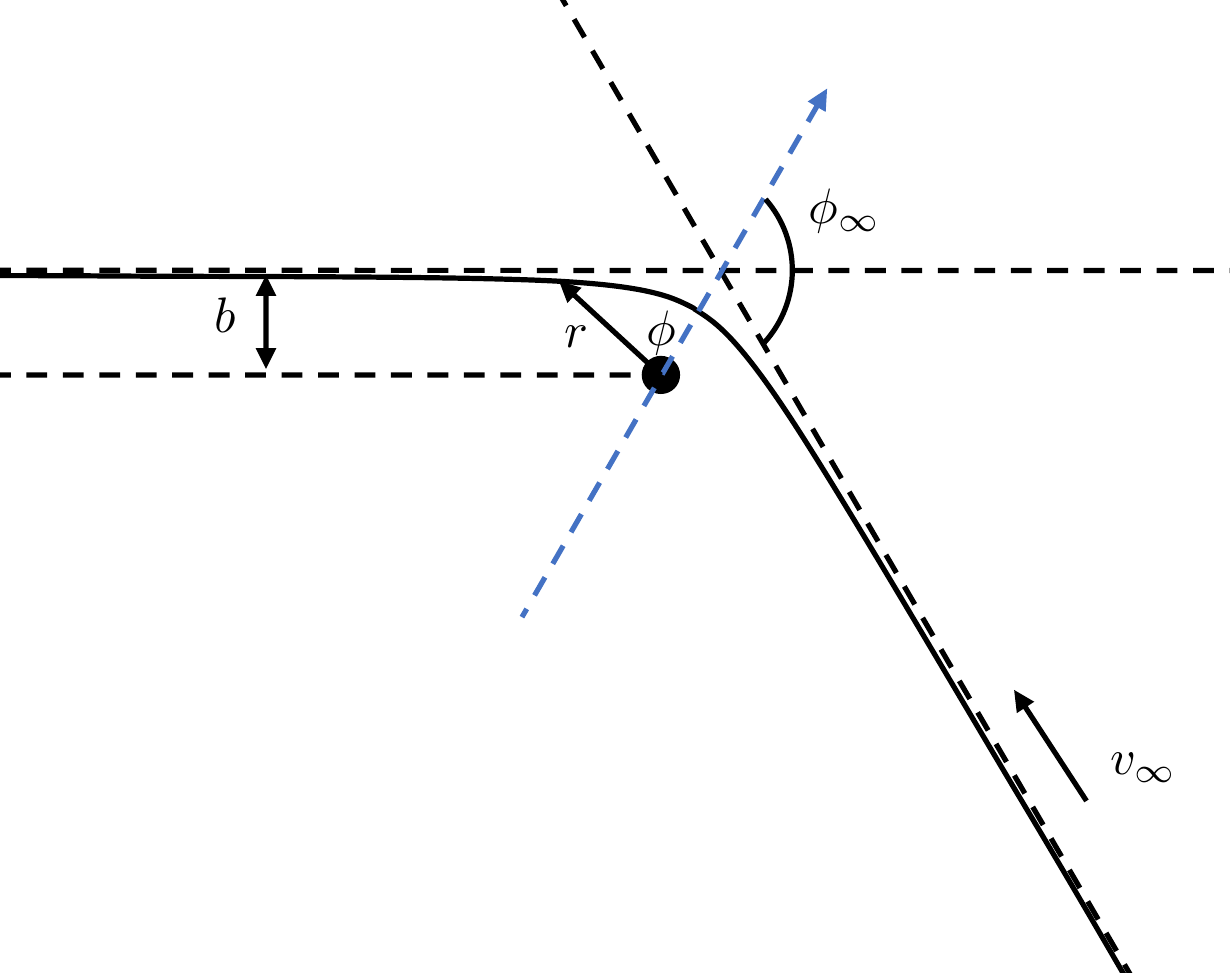}
    \vspace*{-0.5cm}
    \caption{The geometry of the system, where $b$ is the impact parameter 
    and $\phi_\infty$ is the angle between the position vector at periapsis and 
    position vector at $t \to \pm \infty$. 
    When this curve represents geodesic motion, 
    the impact parameter is a constant before and after the scattering.
    The scattering angle is also symmetric.
    }
    \label{fig:geometry}
\end{figure}


When a test object has $E > 1$, it follows a hyperbolic orbit around 
  the central massive \ac{bh}, in which its high angular momentum prevents 
  it from being captured. 
Fig.~\ref{fig:geometry} shows the geometry of such a system, 
  when the orbit is confined to the equatorial plane. 
The equatorial hyperbolic orbit is symmetric with respect 
  to the periapsis (i.e. $\phi_{- \infty}=-\phi_{+\infty}$), when spin's effects are ignored. 
The deflection angle is defined as $\Delta \phi = 2\phi_{\infty} - \pi$, 
  where $\phi_\infty$ is shown in Fig.~\ref{fig:geometry}. 
With the help of conserved values ($E$ and $L_z$),
  the hyperbolic-like orbit of a spinning object scattered by a Schwarzschild \ac{bh} \citep{Bini2017a}
  or a Kerr \ac{bh} \citep{Bini2017b} can be described analytically up to linear order in spin. 
The azimuthal angle at infinity takes the form: 
  \begin{equation}
  \begin{aligned} \label{eq:PhiInfinityAnalytic}
  \phi_{\infty} & = \phi_0(\chi_{\rm max}) + \hat{s} \phi_{\hat{s}}(\chi_{\rm max}) \ , \\
  \end{aligned}
  \end{equation}
where $\phi_0(\chi)$ and $\phi_{\hat{s}}(\chi)$ are defined in Eq.(34) of \citet{Bini2017b}. 
Note that since the analytical formula that is quadratic in spin is 
  not available,
  we compare our numerical results (which is quadratic in spin) with Bini's formula 
  at linear order to perform a consistency check.

\begin{figure}
  \includegraphics[width=1\columnwidth]{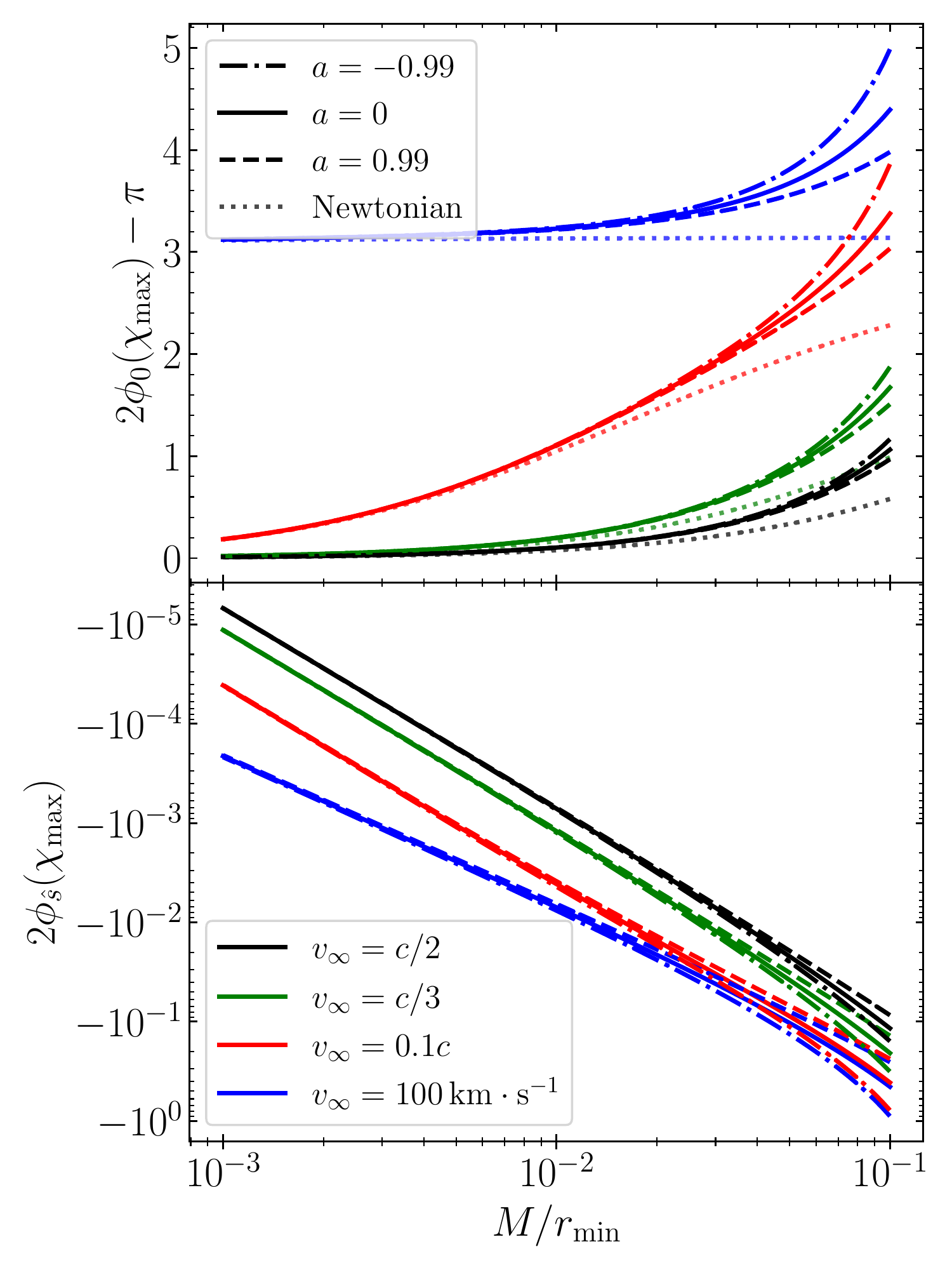}
  \vspace{-0.7cm} 
  \caption{ 
The deflection angle for a time-like test object on geodesic motion (upper panel) 
  and the linear order correction due to spin (lower panel). 
In the upper (lower) panel, the solid line from top to bottom (bottom to top) corresponds to test objects with 
  $E_0 \equiv 1/\sqrt{1- v_{\infty}^2}$ where $v_{\infty}  = 100 \kms$, $0.1c$, $c/3$ and $c/2$, respectively. 
The angular momentum $J_0$ is determined by $v_{\infty}$, $r_{\min}/M$ and $a/M$.
The dashed line and dash-dotted line represent $a/M=0.99$ and $a/M=-0.99$, respectively. 
    }
    \label{fig:DemoDeflectionAngle}
\end{figure}

An example of the deflection angle is shown in Fig.\ref{fig:DemoDeflectionAngle}.
Particles with larger $v_\infty$ have larger inertia, 
  and therefore are more difficult to be deflected. 
They follow orbits with much larger eccentricity. 
Examples of such particles include photons, neutrinos and high energy electrons. 

For stars with $v_\infty \ll 1$, the unit energy of the test object and the orbit eccentricity
  are both close to $1$. 
The eccentricity for such orbits is approximately (exact for Newtonian cases)
  \begin{equation}
  \begin{aligned} \label{eq:SmallEccentricity}
  e \approx 1+\frac{r_{\min}}{M} \frac{v_{\infty}^{2}}{c^{2}} \ , 
  \end{aligned}
  \end{equation}
when $v_\infty \le 100 \kms $ (for any $v_\infty$ in Newtonian cases).  
For such small eccentricity, 
  Newtonian gravity predicts a $\Delta \phi \approx \pi - 2\sqrt{2 (e-1)}$, 
   assembling that of an elongated ellipse with infinity semi-major axis. 
General relativistic effects, in general, increase the deflection angle in the form of 
  periapsis advance. 
Both the relativistic correction and the spin correction to the deflection angle
  are greater for smaller eccentricity, 
  making slow speed objects better laboratories for testing the general relativistic effects.

Currently, the formula for the deflection angle of a spinning test object around 
  a Kerr \ac{bh} is derived for cases where the spin 
  is aligned (or anti-aligned) with the spin of \ac{bh} and orbital angular momentum, 
  in which cases, the orbits are confined to the equatorial plane. 
For non-equatorial orbits, the formula for the deflection angle
  even for geodesic orbits remains unclear.

When comparing an analytic prediction with numerical simulation results, it is impossible to integrate the orbit to infinity
  for the \ac{mpd} equations, 
  and in practice different methods are used to bypass this problem. 
For example, \citet{Damour2014} compare the results at around $r_{\rm f} \approx r_0 =  10^4 M$. 
In this study, we integrate the \ac{mpd} equations of motion from $r_0 = 10^4 M \approx r_{\rm f}$, 
  and integrate geodesic equations from $r_{-\infty}$ to $r_0$ and then $r_{\rm f}$ to $r_{+\infty}$, assuming 
  the effects of spin become negligible in weak field. 
Note that we compare orbits for spinning and non-spinning test objects with the same $E_0$ and $J_0$
  at past infinity.
Therefore, the spin's correction to the deflection angle is not simply $2 \hat{s}  \phi_{\hat{s}}(\chi_{\rm max})$,
  but rather defined as:
\begin{equation}
\begin{aligned} \label{eq:phispinA}
\Delta \phi_{\rm spin, A} & = 2 \phi_0(e', p') + 2 \hat{s} \phi_{\hat{s}} ( e', p') - 2 \phi_0( e, p)  \ , 
\end{aligned}
\end{equation}
where $e$ and $p$ correspond to the geodesic for given set of $E_0$ and $J_0$, 
  and $e'$ and $p'$ differ from $e$ and $p$ at linear order in spin. 
The relation can be found in Appendix.~\ref{ap:AnalyticalFormula}.

The geodesic equations can be written in an alternative form for particles confined to the equatorial plane: 
\begin{equation}
\begin{aligned} \label{eq:PhiInfNumerical0}
\phi_{\pm \infty} & = \phi_{\rm i} \pm \int_{r_{\rm i}}^{\infty} \frac{ \Phi_r(r) + L_z - a E}{\sqrt{R(r)}}  {\rm d} r  {\rm , \, with} \\
R(r) & = \left( E \left( r^{2}+a^{2} \right) - a L_z \right)^{2}-\Delta \left(r^{2}+(L_{z}-aE)^{2}+Q\right) \ , \\ 
\Phi_r(r)  & =  {\frac {a}{\Delta }} \bigg(E \left( r^{2}+a^{2} \right) - a L_z \bigg)\ , \\ 
\end{aligned}
\end{equation}
where the upper (lower) sign is for particles that are moving to (from) infinity. 
$\phi_{\rm i}$ and $r_{\rm i}$ are ``initial condition'' of the integrand,
  which are $\phi_{\rm f}$ and $r_{\rm f}$ for $\phi_{+\infty}$, 
  $\phi_0$ and $r_0$ for $\phi_{-\infty}$. 

When the spin of \ac{msp} is not aligned or anti-aligned with the orbital angular momentum, 
  the \ac{msp} is not restricted to move in the equatorial plane,
  the following alternative form of geodesic equations are used to evaluate $\theta_{\pm \infty}$:
\begin{equation}
\begin{aligned} \label{eq:IntegrationTheta}
& \int_{r_{\rm i}}^{\infty} \frac{{\rm d}r }{\sqrt{R(r)}}  = \pm \int_{\cos \theta_{\rm i}}^{\cos \theta_{\infty}} \frac{{\rm d} (\cos \theta)}{\sqrt{\Theta(\cos \theta)}} {\rm , \, with}  \\
& \Theta (\cos \theta) = Q - \left( Q+a^2 (1-E^2)+L_z^2 \right) \cos^2 \theta + a^2 \left(1-E^2 \right) \cos^4 \theta \ , \\
\end{aligned}
\end{equation}
where the upper (lower) sign is for ${\rm d}\theta/{\rm d} r <0$ ( ${\rm d}\theta/{\rm d} r > 0$ ),
  and $\theta_{\rm i}$ are the initial condition of the integrand,
  which is $\theta_{\rm f}$ for evaluating $\theta_{+\infty}$, 
  and $\theta_{\rm 0}$ for evaluating $\theta_{-\infty}$.  
For the azimuthal direction, the values at infinities are: 
\begin{equation}
\begin{aligned} \label{eq:IntegrationTheta2}
& \phi_{\pm \infty} = \phi_{\rm i} 
   \pm \int_{r_{\rm i}}^{\infty} \frac{ \Phi_r(r)  - a E}{\sqrt{R(r)}}  {\rm d} r  \\ 
& \,\;\;\quad\,\; - {\rm Sign}\left(\frac{{\rm d} \theta}{{\rm d} \tau} \right) \int_{\cos \theta_{\rm i}}^{\cos \theta_\infty}  \frac{ \Phi_\theta (\cos \theta)}{ \sqrt{\Theta(\cos \theta)} }  {\rm d} (\cos \theta) {\rm , \, with} \\
& \Phi_\theta (\cos \theta) = \frac{ L_z}{ 1 - \cos^2 \theta} \ .  \\
\end{aligned}
\end{equation}
where ${{\rm d} \theta}/{{\rm d} \tau}$ is the value at the ``initial condition''. 
Due to the effect of spin, the hyperbola is no longer symmetric, 
  and therefore $\phi_{- \infty} \ne -\phi_{+ \infty}$, and the deflection angle becomes $\Delta \phi = \phi_{+\infty}-\phi_{-\infty} - \pi$. 
We define $\Delta \phi_{\rm spin} = \Delta \phi_{\lambda =1} - \Delta \phi_{\lambda =0}$ to be the correction to 
  the deflection angle due to spin. 

\subsection{Validation of the numerical solution}

\begin{figure}
  \includegraphics[width=1\columnwidth]{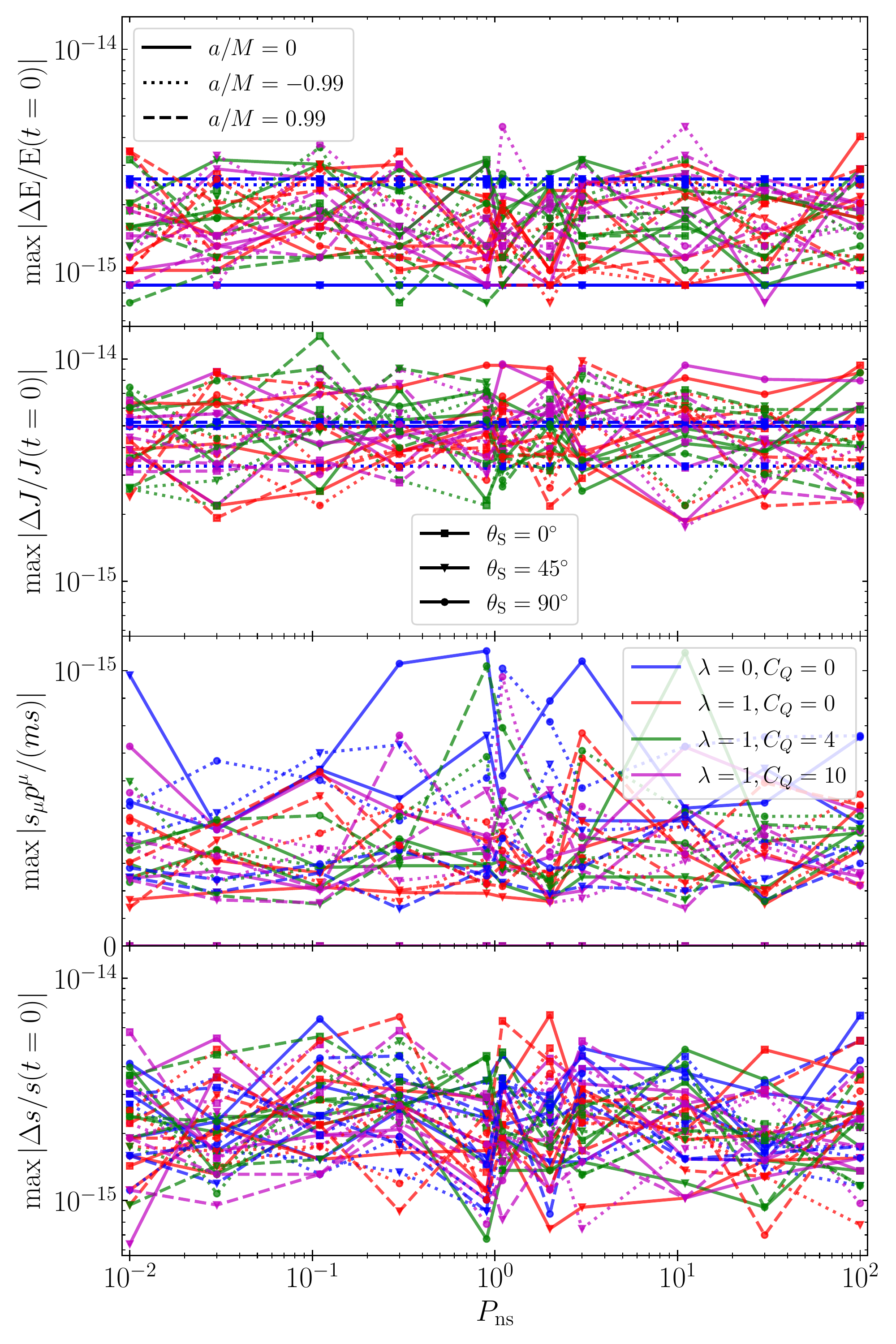}
  \vspace{-0.7cm} 
  \caption{
The system considered here is a (unrealistic) pulsar with period $P_{\rm ns}$, 
  $m=1.5 \solarmass$, $v_{\infty} = 10\kms$ and $J_0 = J_0(r_{\min}=10M)$
  flying by a massive BH 
  with $10^3 \solarmass$ and $a/M= 0, -0.99, 0.99$.
The figure shows 
  the maximum difference of the constants compared with their initial values for $E$,$J$, $s_{\mu} p^{\mu}$ and $s$.
  The value $s_{\mu}p ^{\mu}$ is zero for TD condition and remains less than $2 \times 10^{-5} \times m \times s$ during
    the scattering process.}
    \label{fig:Constants}
\end{figure}

We solve the MPD equation with a 21-stage 10th order Runge-Kutta scheme 
  with an embedded 9th order method to control the time step\footnote{The coefficients are obtained by Peter Stone and are shown in 
  \href{http://www.peterstone.name}{http://www.peterstone.name}.}.
To validate our calculation, 
  the conservation of constants 
  (and semi-constants) are checked 
  for the parameter space of our interest.
Fig.~\ref{fig:Constants} shows the maximum violation of the these constants (and semi-constants) during 
  the evolution from $10^4M \to r_{\min}  \to 10^4M$ 
  with different EOMs (with or without spin-coupling and quadrupole coupling forces)
  for systems with $v_{\infty}= 10 \kms$ and $J_0= J_0 (r_{\min} = 10M)$.
The parameters of the system cover different spin magnitudes ($\hat{s}$ from $5.68 \times 10^{-6}$ to $5.68 \times 10^{-2}$)
  and different orientations of the spin axis.
The energy, angular momentum, spin magnitude and a further constant $s_{\mu}p^{\mu}$ 
  are all conserved with ratio error $ < 2 \times 10^{-14}$ for all three different EOMs.  

\begin{figure}
  \includegraphics[width=1\columnwidth]{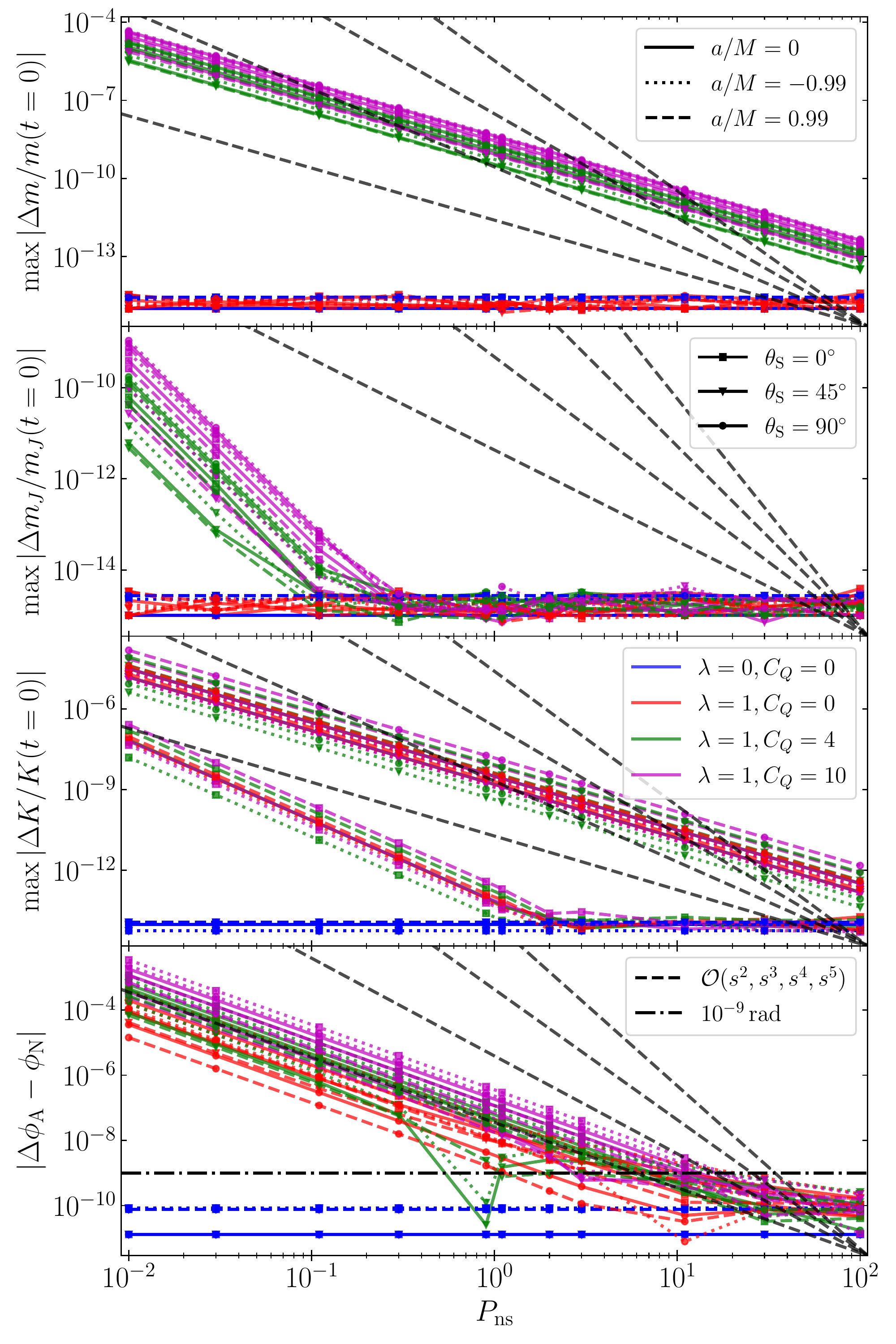}
  \vspace{-0.7cm} 
  \caption{The figure shows 
  the maximum variation of semi-constants: dynamical mass $m \equiv \sqrt{- p^{\mu} p_{\mu}}$,
  the modified mass $m_{J}$, the Rudiger constant $K$ and the difference between numerical 
  deflection angle $\Delta \phi_{\rm N}$ (calculated by Eq.~\ref{eq:PhiInfNumerical0} and Eq.~\ref{eq:IntegrationTheta2})
  and $\Delta \phi_{\rm A} \equiv 2 \phi_0(\chi_{\rm max}) + 2 \hat{s} \phi_{\hat{s}}(\chi_{\rm max}) - \pi$. 
  The black dashed lines in each panel represent the reference lines which are $\propto s^2, s^3, s^4, s^5$.
}
    \label{fig:semiConstants}
\end{figure}

Fig.~\ref{fig:semiConstants} shows the maximum variation for the semi-constants,
  including the Rudiger constant $K$, the modified mass $m_J$ and the dynamical mass $m$. 
The dynamical mass $m$ is a constant when $C_Q=0$ or $\lambda=0$, 
  and vary as $\mathcal{O}(s^2)$ when $C_Q \neq 0$. 
As argued in the previous section, the Rudiger constant is not strictly conserved, 
  but vary as $\mathcal{O}(s^2)$ when the orbit does not follow geodesic, 
  and vary as $\mathcal{O}(s^3)$ when the spin is aligned (or anti-aligned) with 
  the orbital angular momentum. 
The mass-like definition $m_J$ 
  varies as $\mathcal{O}(s^4)$ when $C_Q \neq 0$.

Further, the deflection angles for cases with spin being parallel to the orbital angular momentum 
  are compared for our numerical results with the linear-in-spin analytical formula. 
The spin will introduce an additional shift of periapsis and change the deflection angle at spatial infinity. 
In the lower panel of Fig.~\ref{fig:semiConstants}, the difference in two deflection angles, i.e. $\Delta \phi_{\rm N}$
  and $\Delta \phi_{\rm A}$, are shown for different magnitudes of spin.

When $\lambda =0$ and $C_Q=0$, 
  the two formulae deviate by $< 10^{-9} {\rm rad}$,
  for systems with $v_{\infty} = 10 \kms$ and $J_0 = J_0(r_{\min}=10M)$.
The value $10^{-9} {\rm rad}$ is the limit of accuracy for double precision, 
  for systems with $v_{\infty} = 10 \kms$.
A discussion of this accuracy is included in Appendix \ref{sec:Accuracy}. 
When $\lambda=1$ and $C_Q \neq 0$, 
  the difference comes from a combination of numerical error and second-order spin effects.
  The difference varies as $\mathcal{O}(s^2)$ 
  in the lower panel of Fig.~\ref{fig:semiConstants}.

When $180^\circ > \theta_{\rm S} > 0^\circ$, the \ac{msp} is in fact not confined to the equatorial plane,
      and therefore the analytical formula Eq.~\ref{eq:PhiInfinityAnalytic} are not exact solutions even at linear order. 
Nevertheless, the lines with triangle and circle markers in the lower panel of Fig.~\ref{fig:semiConstants}
  indicate that the analytical formula are still good approximations at linear order.

The variation in deflection angle (as well as periapsis shifts) is mainly due to the force in the radial direction, 
  and therefore is solely contributed by the component of spin that is parallel to orbital angular momentum 
  \citep[see, for example, Eq.~2.2c of][]{Kidder1995}.

\subsection{Motions of MSP at infinity}
\label{subsec:EquatorialMotions}

\begin{figure}
  \centering
  \vspace*{-0cm}
  \includegraphics[width=1\columnwidth]{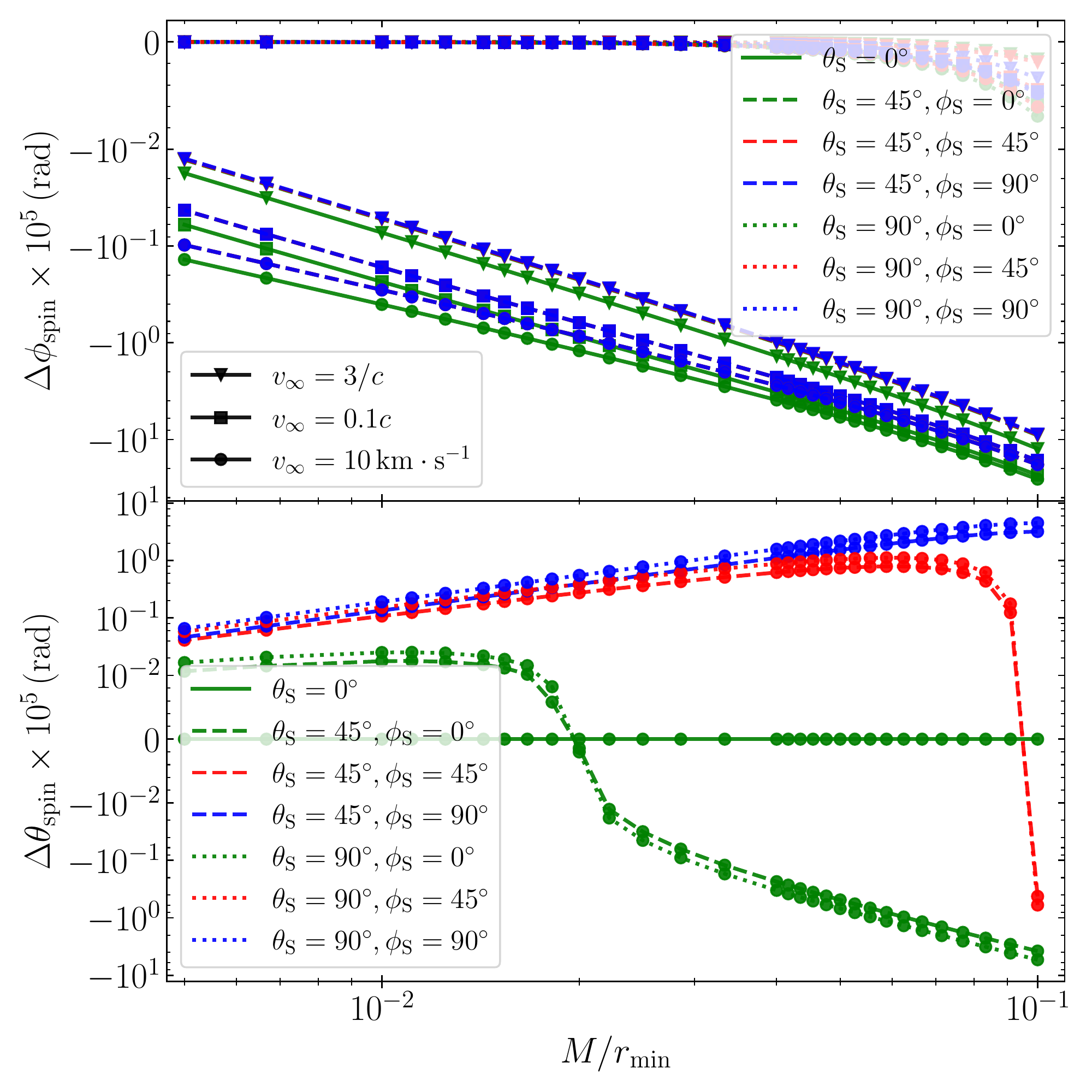}
  \vspace*{-0.7cm}
  \caption{ 
The corrections to the two deflection angles (i.e. $\Delta \phi_{\rm spin}$ and $\Delta \theta_{\rm spin}$)
  due to the spin the of \ac{msp} with different spin orientations.
The spin axes of the \acp{msp} are either perpendicular to the equatorial plane ($\theta_{\rm S}=0^\circ$, solid lines),
  or tilted at $\theta_{\rm S}=45^\circ$ (dashed lines) or $\theta_{\rm S}=90^\circ$ (dotted lines) with 
  respect to the $\hat{z}$.
The BH is non-spinning and has mass $M=10^3 \solarmass$.
The upper panel shows the \ac{msp} with different energies (i.e. $v_{\infty}$) and angular momenta.  
The lower panel shows the \ac{msp} with $v_{\infty} = 10 \kms$ and different angular momenta. 
}
\label{fig:DeflectionFig2}
\end{figure}

Fig.~\ref{fig:DeflectionFig2} shows the deflection angles for systems with different velocities (i.e. different $E_0$),
  different $J_0$ and spin directions.
The effects of spin favour a system with small $v_\infty$ against a system with relativistic speeds.
As shown in the figure, the spin's correction to deflection angle of slow-speed test objects can be larger than
  those of relativistic $v_\infty$ by about $10$ times. 
Combining the information of the scattering cross sections, 
  \acp{msp} with small velocities (e.g. $5-10 \kms$) are favoured against 
  higher speed \acp{msp} as 
  they have smaller inertia and are easier to be scattered; 
  the scattering cross section is much larger (e.g. $b_{v_\infty =10 \kms}/b_{v_\infty =100 \kms} \approx 10$);
  and the effects of spin are more prominent. 
The third point can be visualised from both Fig.~\ref{fig:DemoDeflectionAngle} and Fig.~\ref{fig:DeflectionFig2}.
Smaller $v_\infty \ll$ leads to larger $\phi_{\hat{s}}(\chi_{\max})$ and hence larger $\Delta \phi_{\rm spin,A}$. 

As shown in the upper panel of Fig.~\ref{fig:DeflectionFig2},
  when the \ac{msp} has $v_\infty \ll c$
  and when $\theta_{\rm S}$ is not far from $0^\circ$ or $180^\circ$,
  the deflection angle can be well approximated by
\begin{equation} \label{eq:DeltaPhiSpin}
\Delta \phi_{\rm spin} \approx c_{\phi} \cos \theta_{\rm S} \left( \frac{1\,{\rm ms}}{P_{\rm ns}} \right)
  \left( \frac{10^3 \solarmass}{M} \right)  \ ,
\end{equation}
where $c_{\phi}$ is a function of $r_{\min}$ and $a$. 

\begin{figure}
  \centering
  \vspace*{-0.5cm}
  \includegraphics[width=1\columnwidth]{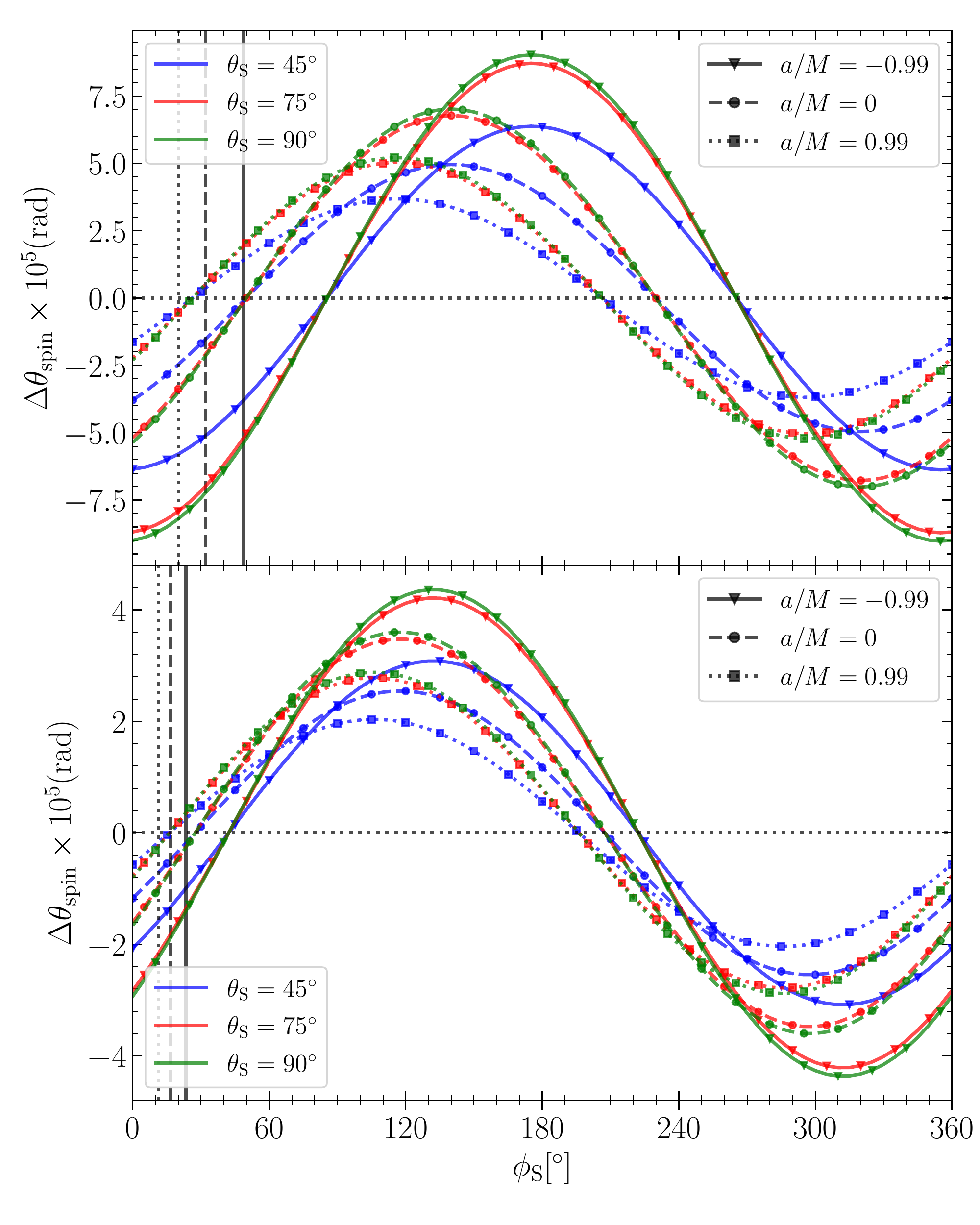}
  \vspace*{-0.7cm}
  \caption{
  The corrections to the deflection angle of the \ac{msp} in $\theta$ direction due to the spin of the \ac{msp}. 
  The \ac{msp} has $v_\infty = 10 \kms$ and $r_{\min}/M=10$ (upper panel) or $r_{\min}/M=15$ (lower panel).
  The BH is non-spinning and has mass $M=10^3 \solarmass$. 
  The vertical lines represent the orientation of the orbit (i.e. $\phi_{\rm per} - \pi$).
  }
  \label{fig:ThetaVariation}
\end{figure}

\begin{figure}
    \centering
    \includegraphics[width=1\columnwidth]{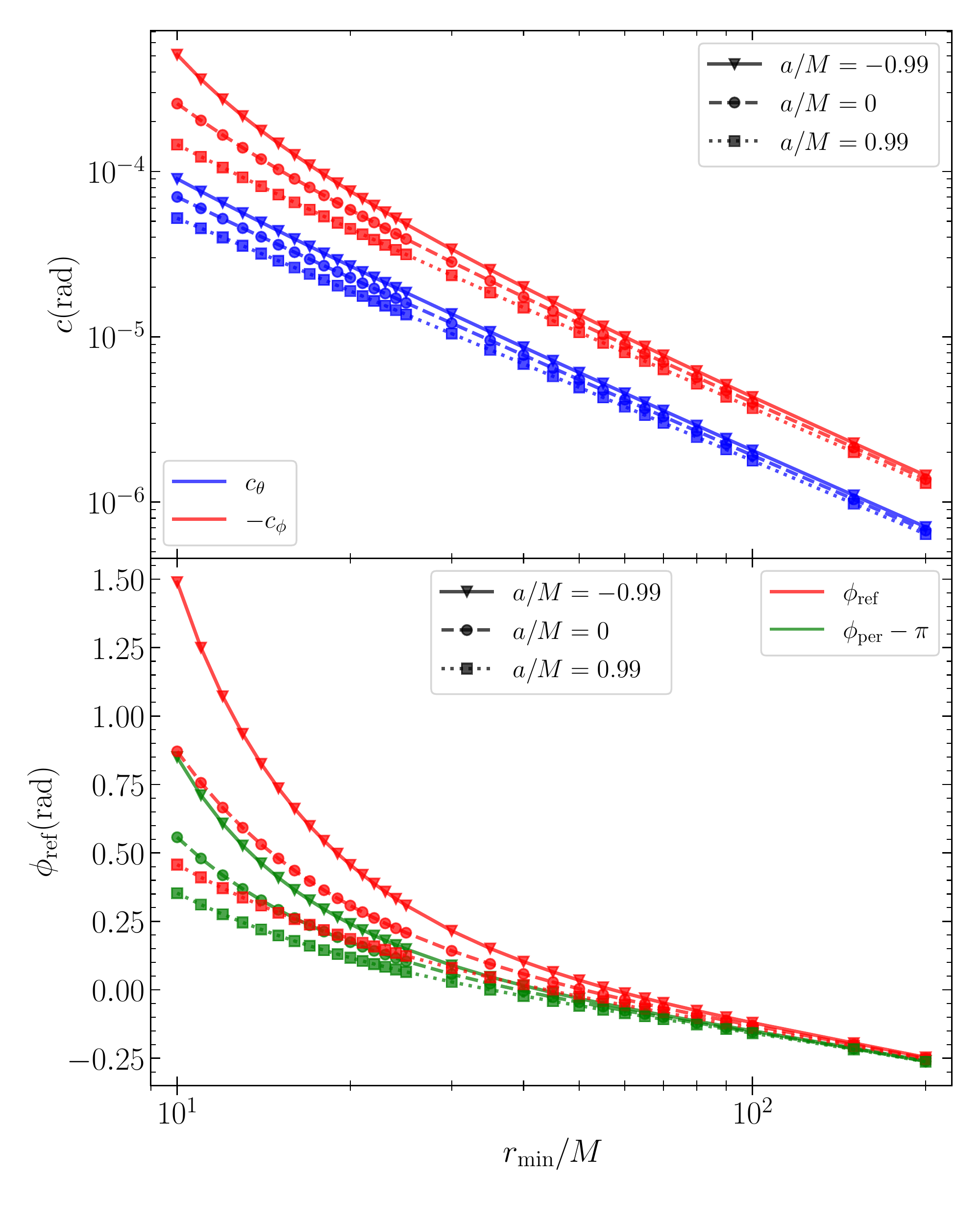}
    \vspace{-0.6cm} 
    \caption{ 
The upper panel shows the value of coefficients $c_{\theta}$ (Eq.~\ref{eq:DeltaThetaSpin})
  and $c_{\phi}$ (Eq.~\ref{eq:DeltaPhiSpin})
  for $10\, M < r_{\min} < 100\, M$ and $a = 0, \pm 0.99 \, M$.
The lower panel shows reference angles $\phi_{\rm ref}$ and the location of periapsis $\phi_{\rm per}$.
}
    \label{fig:KerrDeflectionAnglesFitting}
\end{figure}

When $180^\circ > \theta_{\rm S} > 0^\circ$,
  the orbital plane is tilted around.
The motion of the pulsar is
  usually referred to as the out-of-plane motion.
Such motions have been studied extensively in elliptical orbits \citep[see e.g.][]{Singh2014,Keresztes2019}. 
In hyperbolic scatterings, such out-of-plane motions are also observed. 
The motion in $\theta$ or $z$ direction mainly depends on the components of spin 
  that is perpendicular to orbital angular momentum. 
The coupling between orbital angular momentum and spin allows for the orbit to wobble,  
  and therefore, when \ac{msp} goes to infinity, 
  $\theta_{+\infty}$ deviates from $\theta_{-\infty}=\pi/2$, and such deviation is shown in Fig.~\ref{fig:DeflectionFig2} and Fig.~\ref{fig:ThetaVariation}, 
  in which we define $\Delta \theta_{\rm spin} \equiv \theta_{+\infty, \lambda =1} - \pi/2$,
  which is the correction of deflection angle in $\theta$ direction due to spin. 
This deflection angle for orbits with different parameters is shown in 
  the lower panel of Fig.~\ref{fig:DeflectionFig2} and Fig.~\ref{fig:ThetaVariation}.
The deflection angle can be approximated by:
\begin{equation}
\begin{aligned} \label{eq:DeltaThetaSpin} 
\Delta \theta_{\rm spin} \approx c_{\theta} \sin \theta_{\rm S} \sin \left( \phi_{\rm S} - \phi_{\rm ref} \right)
  \left( \frac{1\,{\rm ms}}{P_{\rm ns}} \right) 
  \left( \frac{10^3 \solarmass}{M} \right)  \ ,
\end{aligned} 
\end{equation}
where $\phi_{\rm ref}$ corresponds to the value of $\phi_{\rm S}$ when $\Delta \theta_{\rm spin}=0$ in Fig.~\ref{fig:ThetaVariation}. 
The coefficient $c_{\theta}$ is also a function of $r_{\min}$ and $a$.
The values of $c_{\theta}$, $c_{\phi}$ and $\phi_{\rm ref}$ are summarised in Fig.~\ref{fig:KerrDeflectionAnglesFitting}. 
The value of $\phi_{\rm ref}$ is related to the the position of periapsis $\phi_{\rm per}$,
  which describes the orientation of the hyperbola, 
  and it gets closer to $\phi_{\rm per} -\pi$ for larger $r_{\min}/M$.  
The slope of $\log(c_{\theta})$ versus $\log(r_{\min}/M)$ and $\log(c_{\phi})$ versus $\log(r_{\min}/M)$
  are all smaller than $-1.4$, implying that this effect is highly relativistic, and is important only
  for very close fly-bys.

When the central BH is spinning, 
  the spin of the BH will introduce an additional shift to the periapsis and $\phi_{\pm \infty}$.
Even a moderate spin of the BH (e.g. $a \approx 0.5 M $) would greatly exceed the spin of the MSP 
  and therefore the spin-orbit coupling will be dominated by the spin of the BH. 
When the BH's spin is in the same direction with the angular momentum, 
  the spin-orbit coupling would introduce an additional force in the opposite direction 
  of Newtonian gravity \citep[see, e.g. Eq.~2.2c of][]{Kidder1995}, 
  preventing the test object from getting too close to the BH, and 
  therefore reduces the deflection angle.  
As readily shown in Fig.~\ref{fig:ThetaVariation}, and Fig.~\ref{fig:DemoDeflectionAngle},
  the retrograde motion has larger deflection angle compared with a comparable prograde motion.

Because the spin of the BH dominates, the effects of the MSP's spin can be seen as a perturbation 
  to the geodesic around the Kerr BH.
Therefore,   
  the effects of the MSP's spin are still approximately symmetrical for $\theta_{\rm S}$ and $180^\circ-\theta_{\rm S}$. 
In fact, both $\Delta \theta_{\rm spin}$ and $\Delta \phi_{\rm spin}$ follow the same relation as Eq.~\ref{eq:DeltaPhiSpin} 
  and Eq.~\ref{eq:DeltaThetaSpin}.
The deflection angles in both directions are the largest for retrograde motion, 
  and smallest for prograde motion.

\subsection{Detectability of the deviation from geodesic}

\begin{figure}
  \centering
  \vspace*{-0.5cm}
  \includegraphics[width=1\columnwidth]{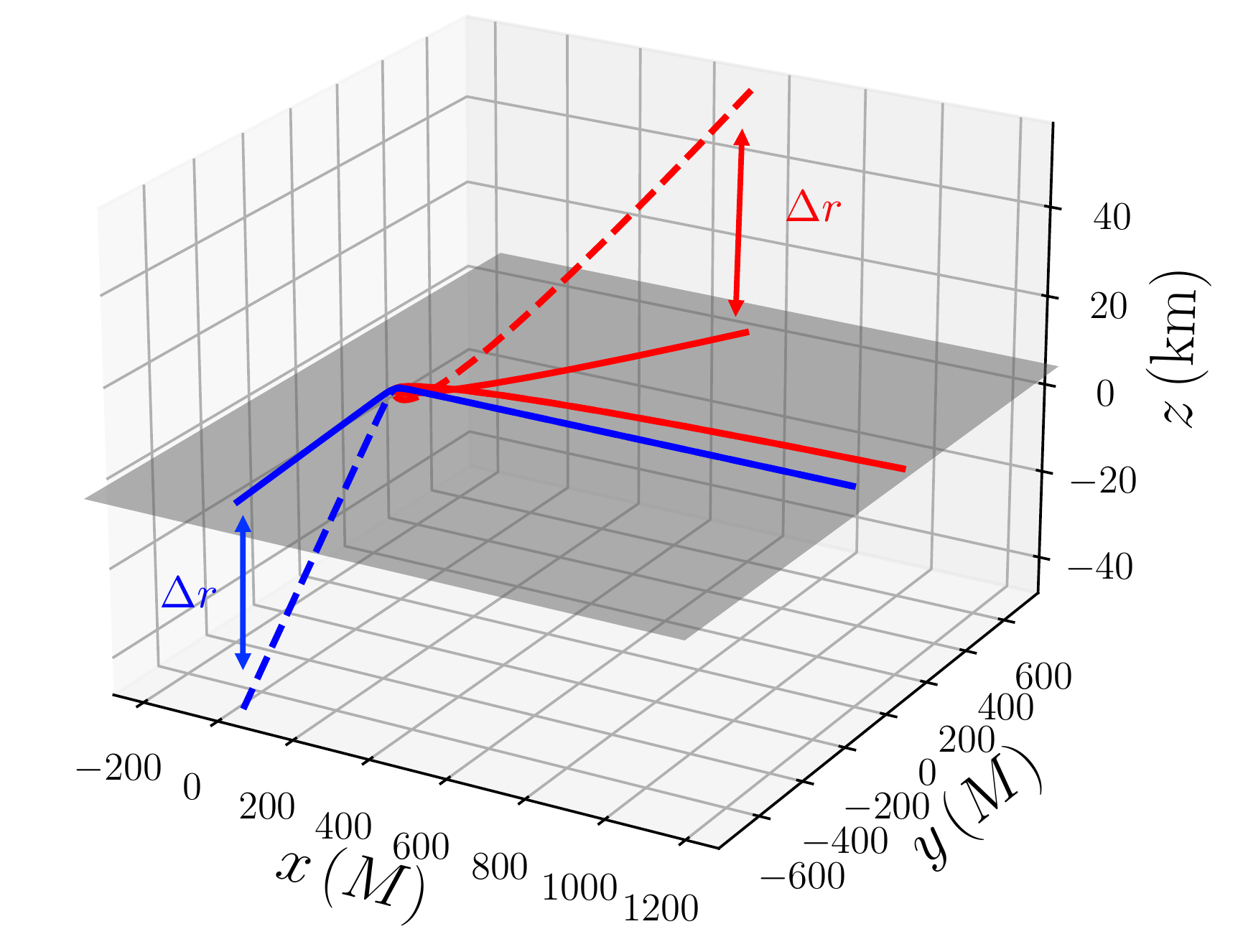}    
  \includegraphics[width=1\columnwidth]{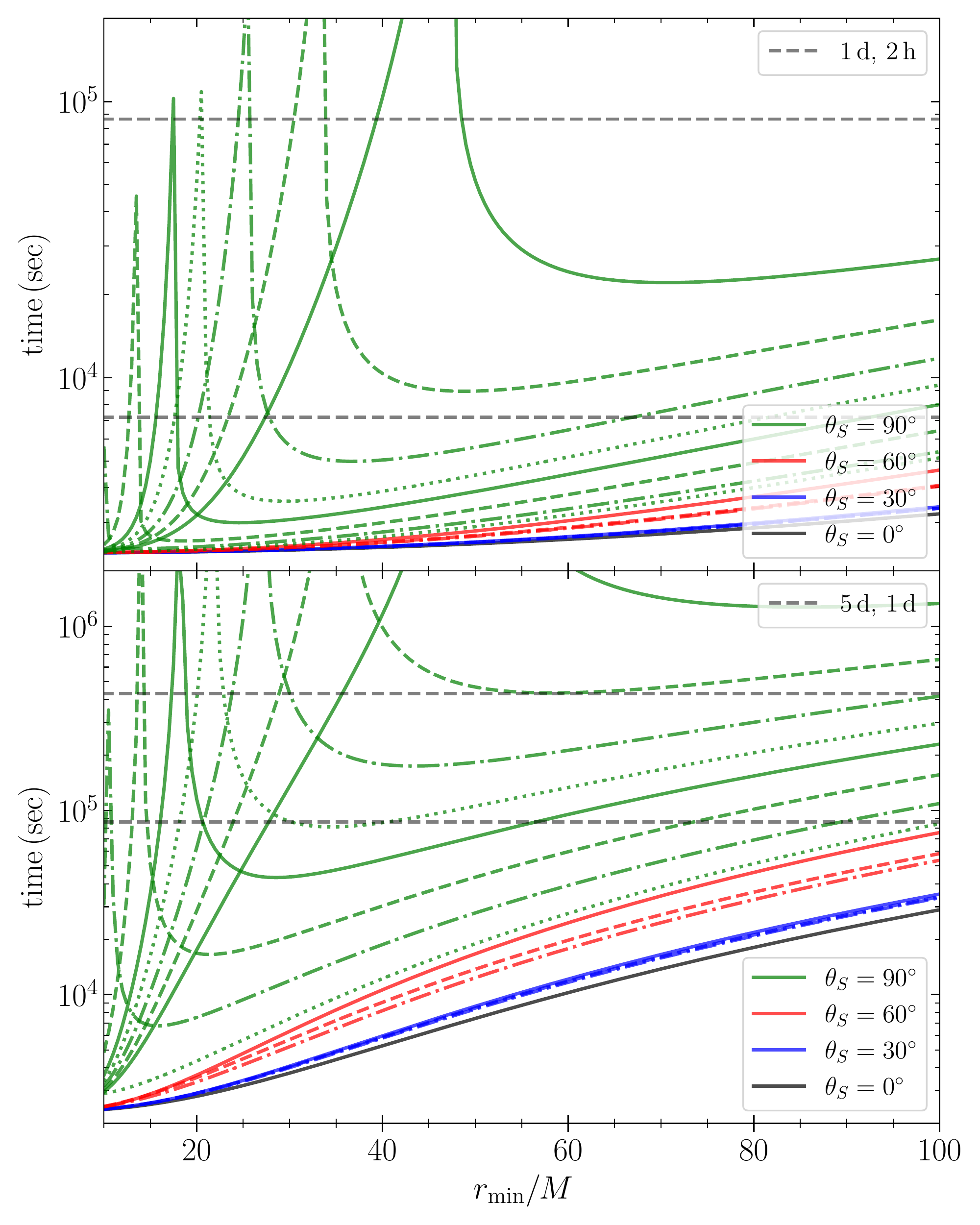}    
  \vspace*{-0.5cm}
  \caption{ 
  This figure shows the observation time it takes for an MSP to deviate from its geodesic orbit
    by $30 {\rm km}$, if the MSP is found at $r=10^4\,M$ approaching the \ac{bh}.  
  The upper panel is an illustration of the deviation. 
 The red and blue lines are geodesic orbits of different energies and angular momenta,
  and the dashed lines are the corresponding non-geodesic orbits. 
  The middle panel shows an MSP pulsar ($P_{\rm ns}=1\,{\rm ms}$).
  The lower panel shows a pulsar with period of $10\, {\rm ms}$. 
  In both the middle and the lower panels, 
   from top to bottom, the first 8 green lines represent \acp{msp} with $\theta_{\rm S}=90^\circ$
  and $\phi_{\rm S}=0^\circ$ (solid) , $5^\circ$ (dashed), $10^\circ$ (dashed-dotted), $15^\circ$ (dotted), 
  $20^\circ$ (solid) , $30^\circ$ (dashed), $45^\circ$ (dashed-dotted), and $90^\circ$ (dotted).
  The successive 3 red lines represent \acp{msp} with $\theta_{\rm S}=60^\circ$
  and $\phi_{\rm S}=0^\circ$ (solid), $45^\circ$ (dashed) and $90^\circ$ (dashed-dotted),
  in which the dashed line and dashed-dotted lines almost overlap. 
  The successive 3 blue lines represent \acp{msp} with $\theta_{\rm S}=30^\circ$
  and $\phi_{\rm S}=0^\circ$ (solid), $45^\circ$ (dashed) and $90^\circ$ (dashed-dotted),
  in which all three lines overlap. 
  The black line at the bottom refers to an MSP with $\theta_{\rm S}=0^\circ$.     
  %
  %
  The horizontal dashed lines represent reference lines of $2\, {\rm hours}$ and one day in the middle panel, 
    and one day and $5$ days in the lower panel.  
  For \acp{msp}, most of the observational times required are below $2$ hours for a close fly-by $r_{\rm min} < 100M$.
  For a pulsar with period $10\,{\rm ms}$, the observational times required are mostly below $2$ days. }
  \label{fig:DeviationTime}
\end{figure}

\begin{figure}
    \centering
    \vspace*{-0.5cm}
 \includegraphics[width=1\columnwidth]{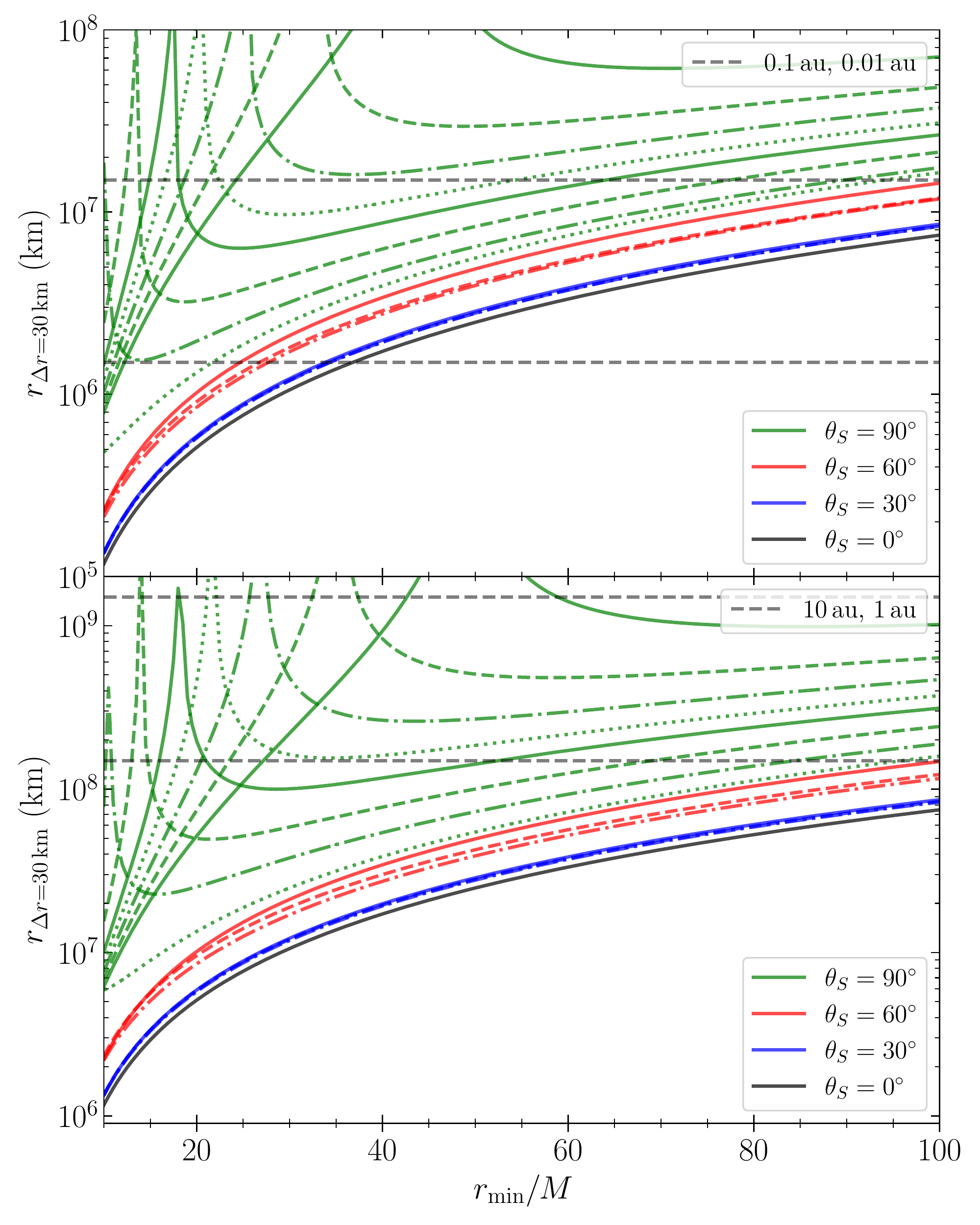}  
    \vspace*{-0.5cm}
    \caption{The radial distance of the MSP with the massive \ac{bh} when $\Delta r = 30\, {\rm km}$. 
    The upper (lower) panel is for a pulsar with spin period $1\,{\rm ms}$ ($10\, {\rm ms}$).
    The parameters of all lines are the same as those in Fig.~\ref{fig:DeviationTime}. 
    The $r$ values are linearly proportional to \ac{bh}'s mass. 
    For both the MSP and the $10\,{\rm ms}$ pulsar, 
      if they undergo a close fly-by ($r_{\rm min}<100\,M$) around the \ac{bh},
      the deviations from the geodesic orbit exceed $30\,{\rm km}$ before reaching $1.4\,{\rm au}$ for most of the spin orientations. 
    }
    \label{fig:DeviationRadialDistance}
\end{figure}

For an $M=10^3 \, \solarmass$ \ac{bh}, a typical deviation in deflection angle is about 
  $\sim 2 \times 10^{-5} \, {\rm rad}$ for a close fly-by ($r_{\rm min}\approx 20M$).  
This deviation in angles corresponds to a spatial deviation of $\sim 30 {\rm km}$ when $r = 1000M$, 
  which could introduce an $0.1 {\rm ms}$ difference in the pulse-arrival-time, different from that of a pulsar following geodesic. 
This is within the precision limit of current pulsar timing technique \citep[see footnote 8 of][]{Li2019}. 
In general, assuming the detectable limit to be $0.1 {\rm ms}$ (i.e. $30 \, {\rm km}$), 
  if a spinning \ac{msp} is found to be flying towards the \ac{bh} at $r= 10^4 M$, 
  the orbit of this \ac{msp} will deviate from the geodesic orbit by $\Delta r = 30 \, {\rm km}$ 
  after a certain observational time. 
This observational time for pulsars with different orbital parameters 
  are shown in Fig.~\ref{fig:DeviationTime}. 
The corresponding radial distance between the \ac{msp} and the \ac{bh} is shown in   
  Fig.~\ref{fig:DeviationRadialDistance}. 

\begin{figure}
    \centering
    \vspace*{-0cm}
    \includegraphics[width=1\columnwidth]{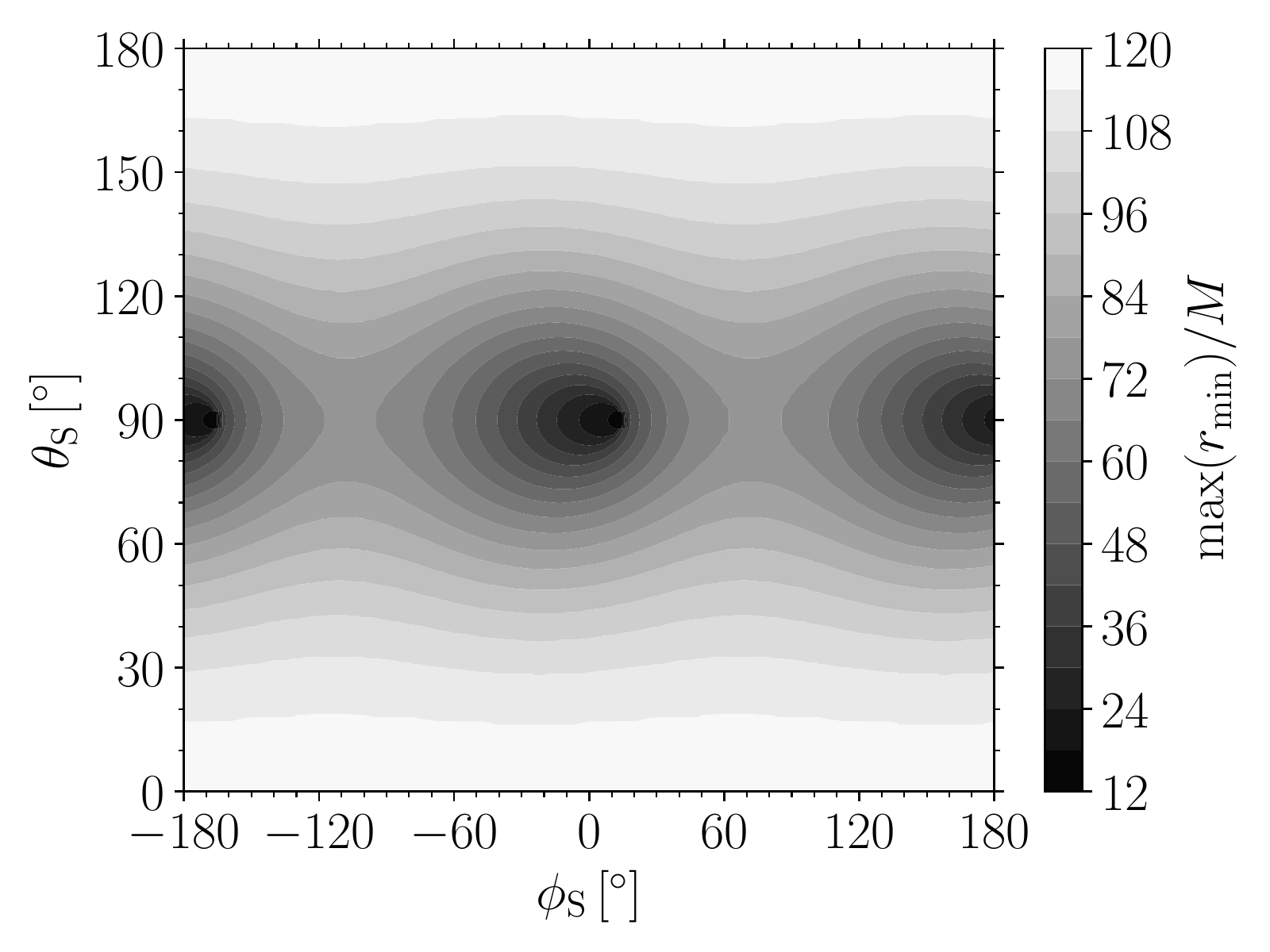}
    \includegraphics[width=1\columnwidth]{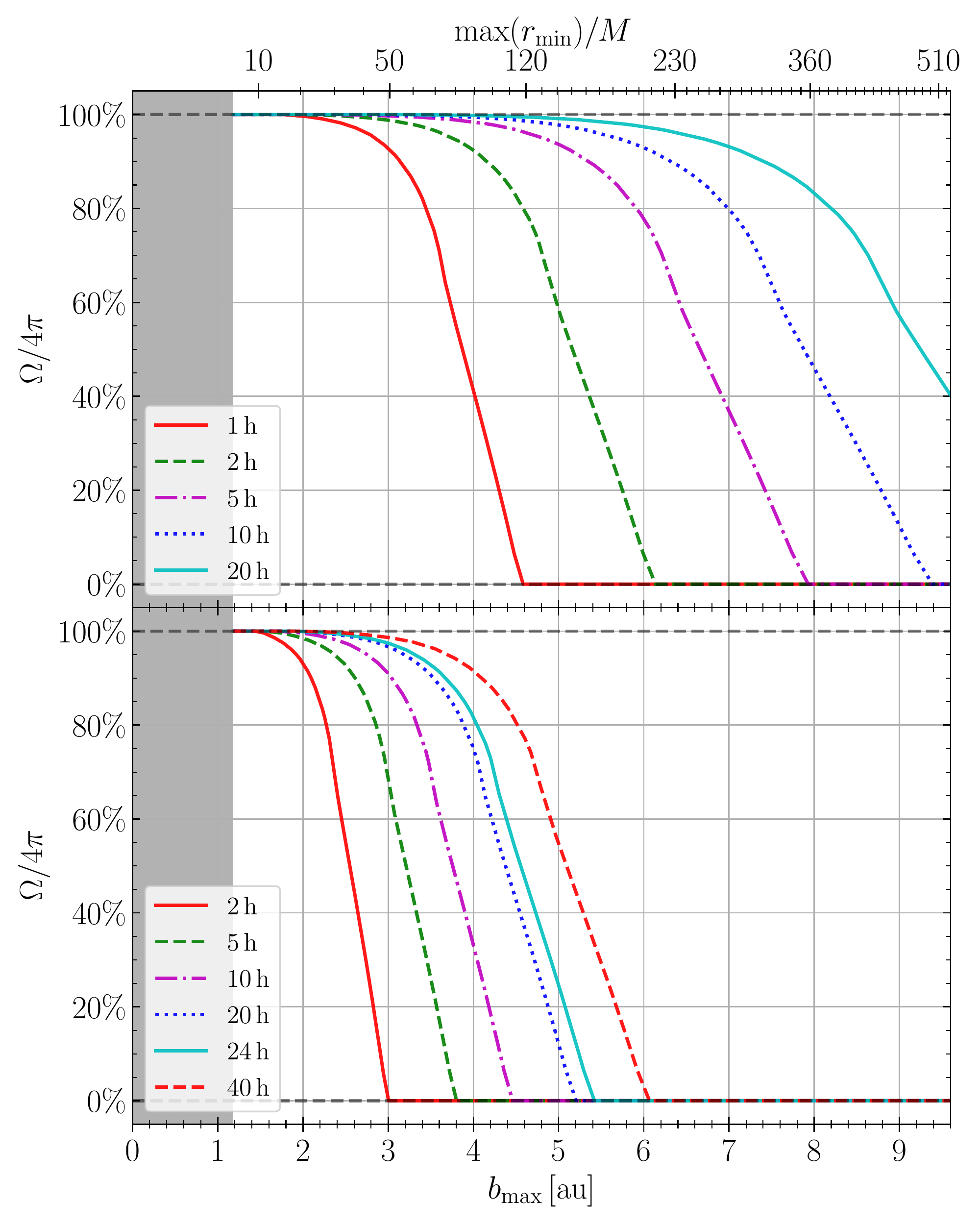}
    \vspace*{-0.5cm}
    \caption{
Suppose that an MSP (with $v_\infty=10\,{\rm km}/{\rm s}^{-1}$) is found at $r=10^4\,M$ 
    moving towards the \ac{bh} with certain orientation 
    (described by $\theta_{\rm S}$ and $\phi_{\rm S}$),
      if the deviation from geodesic is required to reach $\Delta r=30\,{\rm km}$ within 
      1 hour of observation time, there is a maximum required $r_{\rm min}/M$ for this orbit. 
The upper panel shows the maximum required $r_{\rm min}/M$ for different orientations of the spin axis.
The lower two panels show the possibility that a deviation of $\Delta r = 30 \, {\rm km}$
  can be observed 
  within the given observational time, 
  if the orbit has a fixed $r_{\min}/M$,
  for an MSP (middle panel) and a $10\, {\rm ms}$ pulsar (lower panel).  
The possibility is determined by the orientation of the spin axis of the pulsar, 
  and is therefore given in solid angle $\Omega$. 
If $\Omega/4\pi = 1$, 
  the deviation from geodesic can be observed regardless of the orientation of the pulsar,
  while a value smaller than $1$ means that the deviation will be observed only if the pulsar 
  has its spin axis 
  pointing at some preferred directions 
  (in general, parallel and anti-parallel to the orbital angular momentum). 
The shaded regions in lower panels represent the capture domain if gravitational radiation is ignored. 
    }
    \label{fig:DeviationTime1hour}
\end{figure}

If the pulsar is an \ac{msp} that will undergo a close fly-by ($r_{\rm min} < 100\,M$),
  the observational time is in general smaller than $2\,{\rm hours}$ for most orientation of the spin axis. 
If we limit the observational time to be $1\,{\rm hour}$,
  for each possible orientation of the spin axis, 
  there is a maximum $r_{\min}/M$ below which the orbit of the \ac{msp} will deviate 
  from the geodesic by $\Delta r \ge 30\,{\rm km}$ within $1\,{\rm hour}$. 
The maximum value of $r_{\min}/M$ is shown in the upper panel of Fig.~\ref{fig:DeviationTime1hour}. 
It is clear that $\theta_{\rm S} = 0^\circ, 180^\circ$ are favoured orientations (with larger ${\rm max}(r_{\min})/M$),
  and even when $\theta_{\rm S}=90^\circ$, there are still some favoured directions around $\phi_{\rm S} = -110^\circ$ and $ 70^\circ$. 
If an \ac{msp} is found at $r= 10^4 M$ moving towards the \ac{bh} with random orientation, 
  then $\Omega/4\pi=25.8 \% $ of the MSP will have a $0.1\,{\rm ms}$ shift in pulse-arrival-time
  if the pulsar is found to follow an orbit of $r_{\rm min} \le 100\, M$.
Note that $r_{\rm min}\approx 100\, M$ corresponds to a scattering cross-section of $\pi \times ( 4.2\,{\rm au})^2$ 
  for $M = 10^3\, \solarmass$ and $v_\infty = 10\, \kms$. 
  
\begin{figure}
    \centering
    \vspace*{-0cm}
    \includegraphics[width=0.9\columnwidth]{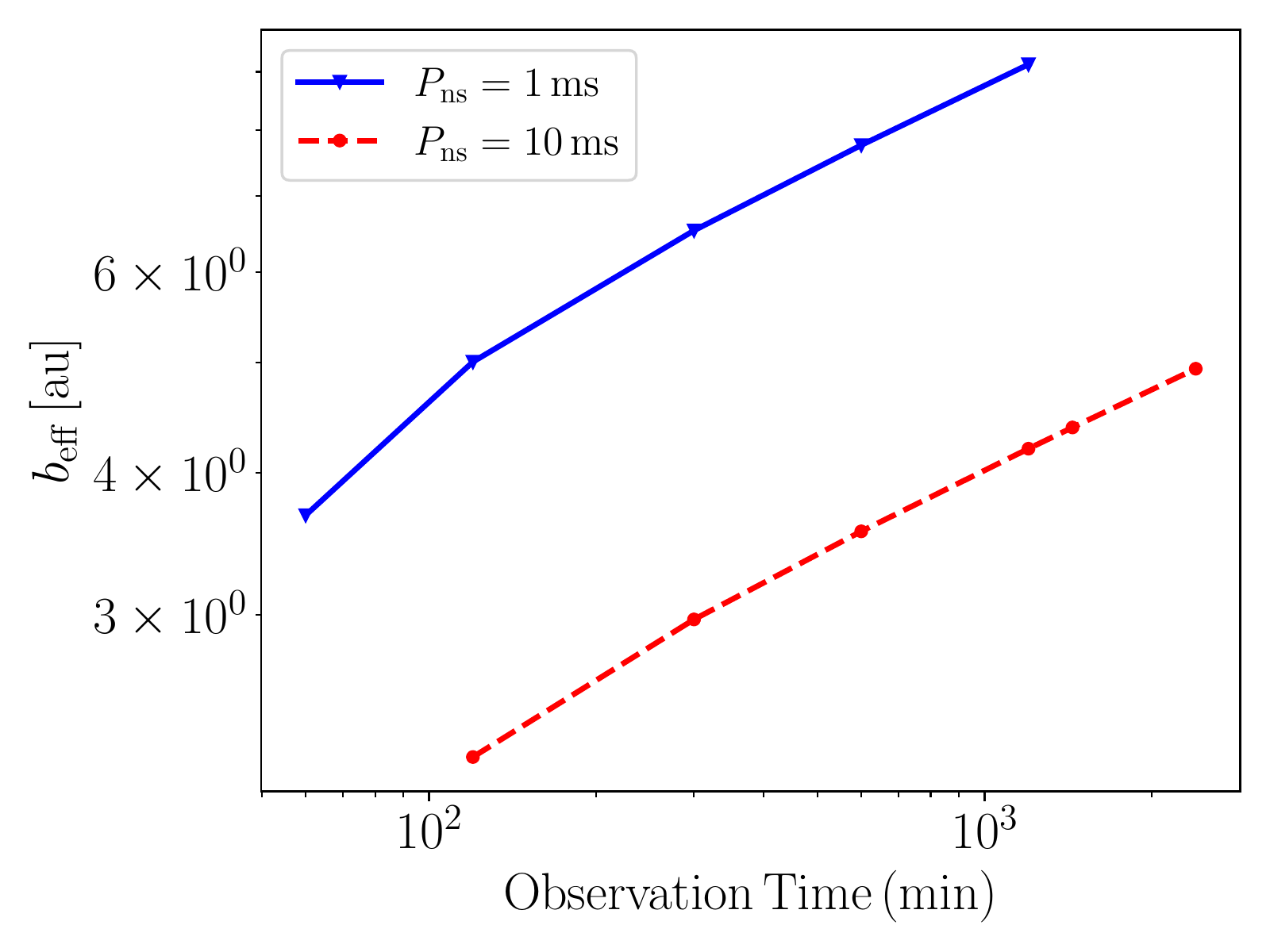}
    \vspace*{-0.5cm}
    \caption{
The effective scattering cross sections $b_{\rm eff}$ for $P_{\rm ns} = 1 \, {\rm ms}$, $10 \, {\rm ms}$,
  and different observation times between $1 \, {\rm h}$ and $40 \, {\rm h}$.
The $b_{\rm eff}$ is defined in Eq.~\ref{eq:EffectiveScattering}.}
    \label{fig:DeviationEffectiveRadius}
\end{figure}

If we suppose \acp{msp} are scattered into the \ac{bh} evenly over some cross section $b_{\rm cs} \le r_{\rm inf}$,
  then the effective cross section $A$, in which the spin-orbit coupling effects can be observed 
  within one hour after the pulsar reaches $r= 10^4 M$ during inward motion, is 
\begin{equation}
\begin{aligned} 
A_{a=0} = & \int_{0}^{b_{\rm cs}} \frac{ \Omega(b)}{4 \pi} 2 \pi b \, {\rm d} b \approx \pi \times (3.7 \, {\rm au})^2 \ . 
\end{aligned} 
\end{equation}  
By setting $b_{\rm cs} \to \infty$, we can define the effective scattering cross section as 
\begin{equation}
\begin{aligned} \label{eq:EffectiveScattering} 
\pi b_{\rm eff}^2 \equiv \int_{0}^{\infty} \frac{ \Omega(b)}{4 \pi} \times 2 \pi b \, {\rm d} b \ . 
\end{aligned} 
\end{equation}  
The effective scattering cross sections for $P_{\rm ns} = 1 \, {\rm ms}$, $10 \, {\rm ms}$, 
  and different observation times are shown in Fig.~\ref{fig:DeviationEffectiveRadius}.

\begin{figure}
  \centering 
  \includegraphics[width=1\columnwidth]{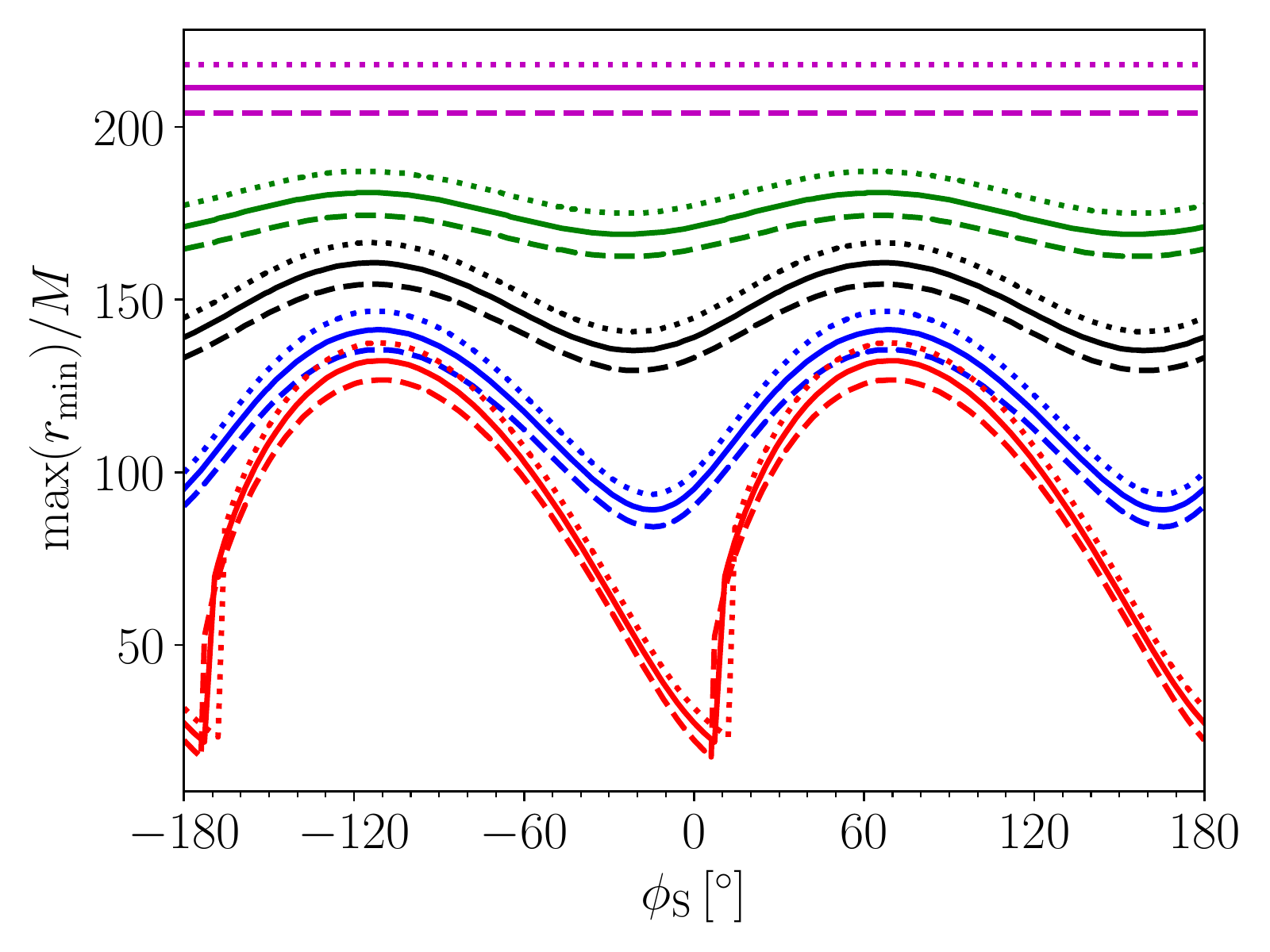}
  \vspace{-0.7cm} 
  \caption{For each assembly of lines, the solid (dashed, dotted) line represent $a=0\,M$ ($a=0.99\,M$, $a=-0.99\,M$). 
  Each bundle of lines from the top to the bottom represents $\theta_{\rm S}$ being $0^\circ$ (magenta), 
  $45^\circ$ (green), $60^\circ$ (black), $75^\circ$ (blue), $90^\circ$ (red), respectively.}
  \label{fig:EquatorialKerrModuleSpinningrMin2D}
\end{figure}

Fig.~\ref{fig:EquatorialKerrModuleSpinningrMin2D} shows the critical $r_{\min}$ for
  the detection of the MSP spin effects 
  if such an MSP fly by a spinning BH, compared with that of a Schwarzschild BH, 
  for MSP with a few typical $\theta_{\rm S}$ values between $0^\circ-90^\circ$. 
In general, prograde motion requires slightly smaller critical $r_{\min}$ while retrograde motion requires slightly larger $r_{\min}$,
  and the locations of the peaks approximately correspond to the values $\phi_{\rm ref} \pm \pi/2$ 
  at corresponding $r_{\min}/M$.

Another measure of the effects of spin-orbit coupling is 
  the radial distance of the pulsar from the \ac{bh} when 
  the deviation from the geodesic reaches $30\,{\rm km}$. 
As shown in Fig.~\ref{fig:DeviationRadialDistance}, 
  the pulsar will deviate from the geodesic by $30\,{\rm km}$ before $r$ reaches 
\begin{equation}
r = 1.4\,{\rm au} \times \left( \frac{M}{10^3 \solarmass} \right) \ ,
\end{equation}  
much smaller than the influence radius for a \ac{bh} between $10^3-4\times 10^6\,\solarmass$. 
This criterion allows us to ignore the effects of surrounding stars of the MSP's orbit and, therefore, 
  simplify the search for such radio pulses.

Because the deflection angle 
  is inversely proportional to the mass of the BH, 
  the radial distance at which deviation from the geodesic reaches $30\,{\rm km}$
  is approximately linearly proportionally to the mass of BH, i.e. $r/M$ is approximately constant. 
The time is, therefore, also linearly proportionally to the mass of BH. 
Hence an observation of one hour for $M=10^3 \solarmass$ corresponds to 
  about $4$ days for $M=10^5 \solarmass$ and 
  about $6$ months for $M=4 \times 10^6 \solarmass$. 
Long time radio observation is possible for important pulsars \citep[see e.g.][]{Weisberg2005,Ransom2020},
  and we do not request the observation data to be continuous in time. 
For the same $r_{\min}/M$, the impact parameter is linearly proportionally to BH's mass. 
Therefore, for more massive BHs, the scattering cross section is much larger ($\propto M^2$). 
If we assume a toy model that the \acp{msp} are scattered into the BHs 
  evenly within the influence radius (which scales as $\propto M^{0.5-0.6}$) with a fixed rate,
  the event rate for more massive BHs is much larger than those of smaller masses, 
  even though the observation of spin effects in scattering by more massive BH requires a much longer observational time. 
A realistic event rate estimate requires detailed modelling of the stellar populations 
  and evolution history of the GC's core or galactic nuclei, 
  and is beyond the scope of this work. 
We look forward to future studies on this issue.

If we allow the detector to track one pulsar for a sufficiently long time (say, e.g. one year), 
  then the detectablity of the spin-orbit coupling is greatly enhanced.
Even \acp{msp} that undergo a distant fly-by will deviate from the geodesic orbit by $30\,{\rm km}$
  before the MSP reaches radial infinity.
Yet it is computationally expensive to calculate these kinds of orbits by solving the \ac{mpd} equations numerically,
  due to the increasing numerical error as we integrate the orbits to large radial distance. 
In other words, a search for even a distant fly-by may require taking spin-orbit coupling into account 
  for long term observations.

\section{Implications in Astrophysics and Physics}
\label{sec:implications} 

\subsection{Presence of pulsars around a massive black hole} 
\label{subsec: pulsars around a massive black hole}

Each massive galactic spheroid, such as an elliptical galaxy or a bulge of a large spiral galaxy, 
  would host at least one massive nuclear \ac{bh}.  
The mass of the nuclear \ac{bh} 
  and the stellar dynamics of the host galaxy are correlated, 
  following an empirical M-$\sigma$ relation \citep{Magorrian1998,Ferrarese2000,Gebhardt2000}, 
  which implies that massive nuclear \acp{bh} reside in massive stellar spheroids.  
Extrapolating the M-$\sigma$ relation 
  down to the lower mass spheroids,  
  we expect that 
  nuclear \acp{bh} of mass $\sim 10^3 - 10^6\;\!{\rm M}_\odot$ 
  would be present in the cores of \acp{gc} and 
  the bulges of spirals and dwarf galaxies 
  \citep[see][]{Miller2002,Reines2015,Mezcua2017}. 
Nuclear \acp{bh} were found in the bulges of many late spirals  
  \citep[see e.g.][]{Jiang2011,She2017}.   
There are also evidences of nuclear \acp{bh}   
   in a number of dwarf galaxies 
   \citep{Reines2013,Moran2014,Nguyen2017},  
   including the compact dwarfs \citep{Voggel2018}.    
The search of nuclear \acp{bh} in \acp{gc} 
  have shown promising results
 \citep{Gerssen2002,Noyola2008,Ibata2009,Feldmeier2013,Lutzgendor2015}, 
  with an intermediate-mass-black hole 
  identified in NGC~6624 \citep{Perera2017} and 47 Tucanae \citep{Kiziltan2017}.  

The dense environment of the \acp{gc} 
  facilitates the formation of \acp{msp} 
  \citep[see e.g.][]{Bhattacharya1991,Manchester2017}. 
The high stellar encounter rates in the core can 
  facilitate the formation of X-ray binaries,   
  where neutron stars can be spun up by the mass-transfer/accretion process \citep{Srinivasan1982,Bhattacharya1991}.    
Indeed, increasing number of \acp{msp} are being found \citep{Lyne1987,Manchester1991,Fermi2013,Freire2017,Dai2020},
  and up to now, 232 \acp{msp} are found among 36 GCs
  \footnote{A list of pulsars in GCs can be found in \\  \href{http://www.naic.edu/~pfreire/GCpsr.html}{http://www.naic.edu/$\sim$pfreire/GCpsr.html}.}.
Population estimation suggests that the actual population of \acp{msp} should be much larger. 
Constraints from gamma ray observations suggest that 2600-4700 \acp{msp} could reside in Galactic GCs, 
  while radio flux observations suggest that 500-2000 \acp{msp} in Galactic GCs 
  \citep{Fruchter2000}, or 12-815 \acp{msp} per GC \citep[see][]{Bagchi2011}.
X-ray observations suggest that ~700 \acp{msp} could reside in Galactic GCs \citep[see][]{Heinke2005}.
Large uncertainty exists even for the prediction of an MSP population in a single GC, 
  due to the complicated dynamical evolution history of GCs.
Some studies try to find a relation between the MSP population with the 
  stellar encounter rate \citep{Bagchi2011,deMenezes2019} or the  
  metalicity \citep{Ivanova2008,Hui2010,Bahramian2013,deMenezes2019} but the populations are preliminarily estimated from gamma-ray or radio observation. 
However, the gamma ray luminosity is not necessarily a good probe of the MSP population, 
  as not all \acp{msp} emit gamma-rays \citep[some might be faint, see e.g.][]{Romani2011,Smith2019},
  and the gamma ray emission in some GCs are dominated by one MSP \citep{Freire2011}. 
Further, for the current radio luminosity model,
  the radio frequency is usually above 400MHz \citep{Bagchi2011,Calore2016}.
A low frequency (115-155 MHz) search indicated that a potential 
  population of MSPs with steep spectrum might have been missed in previous searches \citep{Pleunis2017}. 

 
In our Galaxy, many pulsars are believed to reside in dense nuclei, 
  and \acp{msp} are believed to dominate \citep{Macquart2015},
although no radio observation has confirmed this
  hypothesis, except for a magnetar found at 0.1 pc \citep{Rea2013}. 
The gamma ray excess might be the consequence of this MSP population \citep{Brandt2015,Bartels2016,Fragione2018,Eckner2018}, 
  although some suggest that MSPs could only account for a small portion \citep{Hooper2016}.
The MSP population predicted from different models is highly uncertain, 
  ranging from ~200 \citep[][using the MC method, and over several thousands for Bayesian method]{Chennamangalam2014}, 
  $10^3$ \citep{Wharton2012} to $10^4$ \citep{Rajwade2017} within the inner $1\,{\rm pc}$. 
Nevertheless, the time scale of a close fly-by around Sgr A* is much longer 
  ($\propto M$, and therefore $4000$ times of that of $10^3 \solarmass$ BH).
The scattering cross section will be larger by a factor $10^6$ 
  (because $b \propto M^2$ for same $r_{\min}/M$, but the dispersion velocity is $\approx 100\kms$). 
How this affects the event rate and detectablity of a close MSP-BH fly-by
  requires detailed studies in the future. 

Because of dynamical friction,  
  these \acp{msp} or their progenitors would sink to the bottom of the gravitational well of the stellar spheroids \citep{deMenezes2019}. 
The sinking of the neutron stars to the centre of the stellar spheroid 
  also enhances the chance of encounters of them and the nuclear \ac{bh}, if present, leading to either capture (bounded orbit) 
  or scattering (unbounded orbit).    

Finding \acp{msp} in the core of stellar spheroids has numerous prospects and, therefore, 
  has attracted continuous efforts over decades. 
The \acp{msp} in the Galactic centre could be formed during 
  the local star forming phase \citep[the so-called in situ formation scenario, see e.g.][]{Aharon2015}, 
  or could be the heritage of tidally disrupted GCs \citep{Calore2016,Arca-Sedda2018,Abbate2018}. 
The observations of \acs{msp} populations in the Galactic centre 
  would provide information regarding the formation history of our Galactic nuclei, 
  and constrain the existing astrophysical models of nuclei stellar cluster. 
The observation is challenging,    
  given the complex structures in the interstellar medium 
  at the galactic nuclear region,  
  and is beyond the sensitivity of current surveys \citep{Rajwade2017}. 
The strong scattering and the temporal smearing at low frequencies  
  make it difficult to detect the periodic radio signal 
  \citep{Cordes1997,Macquart2010,Hyman2019}, in particular, 
  for \acp{msp} which are not as luminous as a normal pulsar. 
Not only is the dispersion measure (DM),   
  about $1000\,{\rm pc}\,{\rm cm}^{-3}$ \citep[see e.g.][]{Cordes2002,Yao2017}
  much greater than all GCs in the Galaxy.  
For example, M53, 
  at distance of $\sim 18 \, {\rm kpc}$.  
  much farther away than the Galactic centre, 
  has a DM $\sim 255\, {\rm pc} \, {\rm cm}^{-3}$ \citep{Kulkarni1991}.    
Some studies suggest 
  that the DM in the Galactic centre may also be frequency-dependent 
  \citep{Pennucci2015,Cordes2016}.
Understanding the properties of the interstellar medium 
  in the Galactic centre is crucial for the identifying origins of radio sources 
  and for example, analysing the image of Sgr A* using Event Horizon Telescope. 
Finding a pulsar in the centre of a GC or Galactic nuclei 
  will provide direct measurement of the local gravitational potential. 
For GCs, 
  it allows us to differentiate between different models 
  (e.g. King Model or Plummer model). 
The degeneracy between density distribution and surface luminosity 
  can potentially be resolved by this information direct from the inner core. 
A pulsar in the inner core could also reveal the presence of a BH.
For example, using 3 \acp{msp} in NGC 6752, a large central mass-to-light ratio is found \citep{DAmico2002}, 
  which possibly indicates the existence of a massive BH in the centre. 
The dynamical evolution of GCs depends on the most massive components of the stellar population, i.e. BHs and neutron stars. 
There are believed to be a population of stellar-mass BHs in addition to the neutron stars, 
  as remnants of massive stellar objects \citep[see e.g.][]{Kulkarni1993,Sigurdsson1993}. 
The stellar-mass BH population and \acs{msp} population   
  might not be independent,  
  as both of them sink towards the centre under dynamical friction. 
In fact, an anti-correlation is found 
  for the population of stellar-mass BHs and the population of \acp{msp} 
  if most of \acp{msp} are formed dynamically, 
  because that BHs, if abundant, 
  will dominate the stellar population in the dense core and therefore
  reduce the chance of MSP formation \citep{Ye2019}. 
Observation of \acp{msp} in GCs can put constraints on the stellar-mass \ac{bh} population, 
  which, together with the observation of existing BH population, can provide clues to the early dynamical evolution of GCs. 

The MSP can lose sufficient amount of energy and consequently be captured by the massive BH 
  after a relativistic fly-by ($r_{\min} \le 151\,M$ for $v_{\infty} = 10\kms$ and $M=10^3 \solarmass$, 
  see Sec.~\ref{subsec:physics} for details),
  merging into the BH efficiently, contributing to the growth of a nuclear BH. 
Even one detection of a pulsar-BH scattering event (or null detection) 
  could readily place an lower bound (or upper bound) on the event rate of such a binary. 
This could provide us clues about the 
  formation and growth history of the nuclear BH. 
In addition, precision testing of GR can be performed using pulsar systems:  
  e.g. the test of orbital evolution driven 
  by gravitational radiation loss 
  using the binary pulsar PSR B1913+16 \citep{Weisberg2005}, 
  and the test of the strong equivalence principle 
  using a triple binary system \citep{Voisin2020}.  
The discovery of a pulsar can offer precise measurements of the BH's mass and spin. 
Further, a pulsar-BH system can serve as a perfect laboratory for a variety of tests of GR,
  including, for example, Lense-Thirring precession \citep{Wex1999}, the 
  no-hair theorem \citep{Liu2012,Psaltis2016},
  and signatures to support alternative gravity theories \citep{Liu2014}. 
A pulsar can tell the nature of the massive dark object in the core of GCs or Galactic nuclei,
  whether is it a BH or not \citep{Saxton2016}.


\subsection{Multi-messenger astrophysics}
\label{subsec:GW} 

Currently, radio observation is almost the only channel for identification of \acp{msp}\footnote{Out 
  of 127 \acp{msp} listed in \href{http://tinyurl.com/fermipulsars}{http://tinyurl.com/fermipulsars}, 
  only 10 were not first identified with radio observations. }.
Presumably, such hyperbolic systems are present in the core of GCs or nuclear regions of galaxies,
  and searching for such MSP is made possible by deep radio search 
  targeting at the GCs or nuclear region of nearby galaxies. 
Searching for such MSPs, however, is limited by several factors,
  including the complicated orbital behaviours and observational difficulties.  
In a classical point of view, the motion of the MSP and the radio signal are affected by 
  a variety of effects. 
In the far field (before and after the scattering), the MSP's motion can be approximated with Newtonian theory.   
As the MSP moves close to the BH, the velocity increases approximately as $\sqrt{M/r}$, 
  and hence the Doppler shift becomes important. 
The gravitational redshift factor ($\propto M/r$) also needs to be taken into consideration.
The radio emission from  the MSP is also lensed 
  by the gravitational potential of the BH  
  (and also the surrounding stellar components), 
  casting an additional time delay (i.e. Shapiro delay). 
A proper treatment of these components 
  will allow us to search for the MSP when it is far away from the nuclear BH. 
As the MSP becomes more and more relativistic, a full GR treatment is required because of the 
  high sensitivity of pulsar timing. 
This requires not only the solution to \ac{mpd} equations, 
  including orbital dynamics and spin dynamics \citep[see e.g.][]{Li2019}, 
  but also the general relativistic radiation transfer of the photons \citep[see e.g.][]{Kimpson2019a}. 
An invalid template will lead to the de-phasing of the template with the signal, and as a consequence, a longer observation
  time can lead to smaller signal-to-noise ratio (SNR), 
  putting an artificial obstacle against finding such a system. 
We note that when the gravity field is steep, the time derivative of the observed period derivative can be so large that
  the signal is no longer periodic \citep[see e.g.][]{Blandford1987,Foster1990}. 
We will use the word ``semi-periodic'' to describe such a radio signal. 

As a special class of EMRB system, 
  the hyperbolic encounter emits a burst of GWs,
  and are ideal sources for multi-messenger astrophysics. 
In fact, GW observations can assist the radio observation by using prior knowledge of 
  the location and parameters of such an system deduced from the GW signal. 
For example, \citet{Kimpson2020a} studied the GW emission from a bounded EMRB system. 
For an unbounded system, we can use the Newtonian approximation from \citet{Capozziello2008} to estimate the 
  characteristic strain of the GW: 
\begin{equation}
    \begin{aligned}
h = & \frac{2}{R} \left\langle\ddot{Q}_{i j} \ddot{Q}^{i j}\right\rangle^{1 / 2} 
\approx 4 \sqrt{\frac{2}{3}} \frac{1}{R}  \frac{m  M }{r_{\min}} \cos ^2 \frac{\Theta }{2}  \sqrt{3 \cos \Theta+4}  \, \\
    \end{aligned}
\end{equation} 
where $\Theta$ is the true anomaly, $Q_{ij}$ is the quadrupole mass tensor and $R$ is the distance between 
  this system and earth. 
The peak value is reached when $\Theta \to 0 $: 
\begin{equation}
    \begin{aligned}
h_{\rm peak} = & 4 \sqrt{\frac{14}{3}} \frac{1}{R}  \frac{ m  M}{r_{\min}} \ , \\
 \approx & 6.2 \times 10^{-19} 
 \bigg(\frac{R}{10 \, {\rm kpc}} \bigg)^{-1}
 \bigg(\frac{m}{1.5 \solarmass} \bigg)^{-1}
  \bigg(\frac{ r_{\min}}{100 M} \bigg)^{-1} \ . 
    \end{aligned}
\end{equation} 
Note that this peak value of the strain is greater than 
  that of a similar EMRI system on circular orbit with radius $r_{\min}$ by a factor of $\sqrt{7/3}$,
  as the velocity at the periapsis is much larger than that of the circular orbit. 
The time scale of this GW emission can be characterised by 
  the full-width at half maximum of the GW amplitude. 
Therefore we have $\Theta \to 1.34$: 
\begin{equation}
    \begin{aligned}
\tau = & 2 \sqrt{\frac{ \alpha^3}{ M}} (e \sinh \epsilon - \epsilon)  \approx  \frac{4 \sqrt{2} r_{\min}^{3/2}}{3 \sqrt{G M}}
\frac{  (\cos \Theta  +2) }{(\cos \Theta +1) }  \tan \frac{\Theta }{2} 
\bigg|_{\Theta \to 1.34} \ , \\
\approx & 13.4 \, {\rm s} \times 
 \bigg(\frac{M}{10^3 \solarmass} \bigg)
  \bigg(\frac{ r_{\min}}{100 M} \bigg)^{3/2} \ , 
    \end{aligned}
\end{equation}  
where $\epsilon$ is the eccentric anomaly, $\alpha$ is the (positive) semi-major axis of the hyperbola with $\alpha = b/\sqrt{e^2 -1}$. 
If an MSP is found on such a hyperbolic orbit with a BH in the galactic centre or GCs in nearby galaxies, 
  it will be a very luminous GW source in the LISA band.
In fact, for appropriate masses of the BH ($10^4-10^5 \solarmass$), 
  this system can be detected up to $10 \, {\rm Mpc}$, 
  and even $\sim 100\, {\rm Mpc}$ for optimal orbital parameters by LISA \citep[see][for the LISA sensitivity curve]{Sathyaprakash2009}. 
Detecting a radio pulsar at such a large distance is certainly very difficult. 
However, the GW detection can be used as a trigger,
  and deep targeted search can be performed by SKA or FAST
  once the GW source is located on the sky map.  
The GW data, together with the gamma-ray/x-ray data can be used as a prior for searching
  of such a semi-periodic radio signal.
Due to the high accuracy of radio timing, 
  the discovery of such a semi-periodic radio signal will greatly enhance the scientific gain from such a GW event.

\subsection{Fate of the hyperbolic encounter}
\label{subsec:physics}   

Stars with slow speed can be easily captured by the 
  central massive \ac{bh} via gravitational radiation. 
The energy lost by the MSP (per unit mass) is \citep{Quinlan1989}
  \footnote{Note that this is the results for a parabolic orbit. A hyperbolic orbit with $v_\infty^2 r_{\min} \ll 1 $ 
  in Newtonian approximation yields the same result. }: 
\begin{equation}
    \begin{aligned}
   \Delta E \approx & - \frac{85 \pi \mu M^{5/2} }{12 \sqrt{2} r_{\min}^{7 / 2}} \ . \\
    \end{aligned}
\end{equation}
The maximum value of $r_{\min}$ that leads to 
  the capture is given by \citep{Quinlan1989}:
\begin{equation}
    \begin{aligned}
    r_{\min, {\rm c}} \approx & \left[\frac{85 \sqrt{2}   \pi m M^{5 / 2}}{12  v_\infty^{2}}\right]^{2 / 7} \ , \\
    \approx & 151 \, M 
    \left(\frac{\mu}{1.5 \solarmass} \right)^{2/7}
    \left(\frac{M}{10^3 \solarmass} \right)^{-2/7}
    \left(\frac{v_\infty}{10 \kms} \right)^{-4/7} \ ,
    \end{aligned}
\end{equation}
which corresponds to a capture impact parameter: 
\begin{equation}
    \begin{aligned}
    b_{\rm c} \approx 5 \, {\rm au}
    \left(\frac{m}{1.5 \solarmass} \right)^{1/7}
    \left(\frac{M}{10^3 \solarmass} \right)^{6/7}
    \left(\frac{v_\infty}{10 \kms} \right)^{-9/7} \ .
    \end{aligned}
\end{equation}
This capture impact parameter is comparable 
  to Eq.~\ref{eq:impactparameter},
  and is almost directly proportional to the the mass of the central \ac{bh}. 
Thus, a more massive \ac{bh}, which have a larger capture cross-section,
  will give more stellar encounter events. 
Even though captured, the energy of these \acp{msp} are still very large $E\approx E_0 + \Delta E \approx 1$ that 
  they follow orbits that are nearly unbounded with semi-major axis: 
\begin{equation}
    \begin{aligned}
a \approx  2\times 10^3 \, {\rm au} 
    \left(\frac{\mu}{1.5\,\solarmass} \right)^{-1}
    \bigg(\frac{M}{10^3 \solarmass} \bigg)^{2}
    \bigg(\frac{r_{\min}}{100 M} \bigg)^{2/7} \ ,
    \end{aligned}
\end{equation}
and period 
\begin{equation}
    \begin{aligned}
    P \approx 3 \times 10^3 {\rm year}
    \left(\frac{m}{1.5 \solarmass} \right)^{-3/2}
    \left(\frac{M}{10^3 \solarmass} \right)^{5/2}
    \bigg(\frac{r_{\min}}{100 M} \bigg)^{21/4} \ .
    \end{aligned}
\end{equation}
We note that $2 \times 10^3 \, {\rm au} \approx r_{\rm inf}$ for a $10^3 \solarmass$ \ac{bh}
  \footnote{Assuming $\sigma \approx 18 \kms$. The semi-major axis will be much smaller than the influence radius is 
  $\sigma$ is smaller.},
  and therefore this orbit will likely deviate from Keplerian motion at $r \gg r_{\min}$ 
  due to the gravity of surrounding stellar objects. 
This deviation will be more stringent for more massive \ac{bh}.   
For simplicity, we ignore the effects of surrounding stellar objects for the time being. 
Then, for a very close fly-by $r_{\min} \le 25 \, M$ around a $10^3 \solarmass$ \ac{bh},  
  the MSP will return within one year and become a periodic GW and radio source. 
If we allow for a longer observational time of about, say, $20$ years, 
  then a close fly-by with $r_{\min} \le 44 \, M$ around $10^3 \solarmass$ \ac{bh} 
  will be seen by the future generation of radio telescopes. 
Using a Newtonian approximation, the eccentricity of the captured orbit can be deduced from $L_z = \sqrt{a M (1- e^2)}$,
  and the merger time will be about $T \approx 2 P$ with exactly the same 
  dependencies on $m$, $M$ and $r_{\min}$ \citep[using radiation formula from][]{Peters1964},
  meaning that these \acp{msp} will merge into the BH within few orbital periods, regardless of mass of the BH and periapsis distance. 
These \acp{msp} (with $r_{\min} \le r_{\min,{\rm c}}$) are efficient sources for the growth of the \ac{bh}.

If we take $v_\infty $ to be the dispersion velocity satisfying M-$\sigma$ relation, 
  with index $5.1$ from \citet{McConnell2011} \citep[or $4.38$ from][will give a similar result]{Zubovas2019}.  
  then $r_{\min, {\rm c}} \propto M^{-0.4}$.  
For $M=10^5 \solarmass$, \acp{msp} with $r_{\min} > 24\, M$ can remain unbounded, 
  and for $M=4 \times 10^6 \solarmass$, \acp{msp} with $r_{\min} > 6\, M$ remain unbounded 
  (or $r_{\min} > 4 \, M$ if we take $v_\infty \approx 100 \kms$)
  and will mix with the stellar objects in the core.
Unlike a binary BH which can serve as a energy reservoir 
  and give away sufficient energy to the scattered objects to escape the gravity potential \citep[see e.g.][]{Merritt2013}, 
  a single BH approximately preserves the energy of the scattered particle (if ignoring gravitational radiation). 
Energy is lost (instead of gained in case of a binary nuclear BH) by gravitational radiation,
  allowing for the core to gradually contract. 
Therefore, these unbounded pulsars cannot escape the gravity potential of the core, 
  unless if they have already gained sufficient kinetic energy during scattering with nearby stellar objects,
  before being scattering by the nuclear BH (however, there is no reason to believe that these pulsars 
  with high kinetic energy will fly-by the nuclear BH, due to much smaller scattering cross section).
The rarity of \acp{msp} in the outskirt (i.e. beyond half-light radius) of GCs\footnote{Among the 136 \acp{msp} detected in 36 Galactic GCs
with known offsets,  
from \href{http://www.naic.edu/~pfreire/GCpsr.html}{http://www.naic.edu/$\sim$pfreire/GCpsr.html},
 only 10 are outside the half-light radius.
  Note that the observation of MSPs is biased against core because of higher luminosity and steeper gravity potential. 
  Further, as some GCs have only few \acp{msp} identified, this statistics is not necessarily reliable.  } suggest that
  either a binary nuclear BH phase does not dominate the GC evolution 
  \citep[e.g. the PSR J1911-5958A is believed to be scattered onto the current orbit by a binary nuclear BH, see e.g.][]{Colpi2002} , 
  or most pulsars could not gain sufficient kinetic energy during interactions with stellar objects. 
\subsection{Conservative post-Minkowskian treatment of hyperbolic encounters}
\label{sec:SelfForce}

The self-force is one of the most important issues  
  yet to resolve for the LISA 
  to observe extreme-mass-ratio systems. 
While most studies 
  have put focus on systems in elliptical orbits, 
  systems in hyperbolic motions 
  are gaining much attention recently \citep[see e.g.][etc]{Damour2018,Bjerrum-Bohr2018,Bern2019a,Bern2019b,Damour2020}, 
  due to the non-degenerate feature of the scattering angle. 
Extending the MPD equations to include the self-force remains a non-trivial question \citep{Kopeikin2019}. 
In this section, we only estimate the correction to the deflection angle, 
  which are derived using the post-Minkowskian (PM) method.  
For hyperbolic orbits, the self-force corrections enter the scattering angle at second 
  post-Minkowskian order. 
The corrections are given by
\begin{equation}
\chi_{\rm 2PM} = -\frac{3}{8} \pi  \left(5 E_0 ^2-1\right) \left(1-\frac{1}{\sqrt{2 (E_0 -1) \nu +1}}\right) \   
\end{equation} 
 \citep[][]{Damour2018}, 
where $\nu \equiv m M/(m+M)^2$. 
For an MSP scattered by a BH of  $M=10^3 \solarmass$ and $v_{\infty} \ll c$, we have
\begin{equation}
\chi_{\rm 2PM} \approx -3.5 \times 10^{-3}\left( \frac{10^3 \solarmass}{M} \right) 
  \left( \frac{v_{\infty}}{c} \right)^2   {\rm rad}  + \mathcal{O}\left( v_{\infty}^3 /c^3 \right)  \ , 
\end{equation}
and is about $-3.9 \times 10^{-12} \, {\rm rad}$ when $v_{\infty} = 10 \kms$.
The third order post-Minkowskian correction can be found in \citep[][]{Bern2019a,Bern2019b,Bini2020,Damour2020}
\begin{equation}
\begin{aligned}
\chi_{\rm 3PM} = & - \frac{2 \nu \sqrt{E_0^2 -1}}{ 2 (E_0 -1) \nu +1} \Bigg[ 
\frac{2}{3} E_0 \left(14 E_0 ^2+25\right)  \\
& +
\frac{4 \left(4 E_0 ^4-12 E_0 ^2-3\right)}{\sqrt{E_0 ^2-1}} \sinh ^{-1} \sqrt{\frac{E_0 -1}{2}} \Bigg] \\ 
\approx & - 8 \nu \frac{v_{\infty}}{c} + \mathcal{O}\left( \nu, v_{\infty}^3/c^3 \right) \\ 
\approx & - 1.2 \times 10^{-2} \left( \frac{10^3 \solarmass}{M} \right) \left( \frac{v_{\infty}}{c} \right) {\rm rad} \ . 
\end{aligned}
\end{equation}
This correction is about $-4.0 \times 10^{-7} \, {\rm rad}$ when $v_{\infty} = 10 \kms$.

Therefore, for the hyperbolic scatterings that we are interested in, 
  the contribution of the self-force is negligible when $v_{\infty}$ is sufficiently small. 
For the 2PM self-force, the correction is smaller than $10^{-9}  \, {\rm rad}$ when $v_{\infty} \le 100 \kms$. 
For the 3PM self-force, the correction is of comparable order of $c_{\theta}$ and $c_{\phi}$ for $v_{\infty}=10 \kms$,
  when $r_{\min}/M > 100$, and is only negligible if $r_{\min}/M$ is sufficiently small.

\section{Conclusions}
\label{sec: conclusion}

It is believed that a large number of \acp{msp} exist in GCs and Galactic centre, where 
  a massive nuclear BH is believed to reside.
The \acp{msp} sink towards the centre under dynamical friction and undergo violent scattering with nearby stars
  in the dense stellar environment, potentially forming an unbounded system with the nuclear BH. 
In this work, we investigate the motion of such an MSP on a hyperbolic orbit around a massive BH in an astrophysical
  context. 
Due to the small dispersion velocities $\le 100 \kms$ in GCs and Galactic centre,
  the event rate of such scattering is purely dominated by the gravitational focusing: 
  \acp{msp} with smaller $v_\infty$ can easily follow a close fly-by with the nuclear BH. 
In general, the relativistic effects (including spin's effects of the MSP) are the most prominent for 
  slow speed test objects, due to their smaller inertia. 
While most of the existing studies on the scattering angle (e.g. the studies on conservative self-force contribution, as shown in Sec.~\ref{sec:SelfForce}) 
 focus on the high-energy scattering regime, 
the effects of the relativistic effects might have been underestimated for astrophysical objects.

The orbit of the MSP is calculated by solving the quadratic-in-spin accurate \ac{mpd} equations.
We compare the orbits of spinning MSPs (with dimensionless spin $\hat{s} \approx 5.68 \times 10^{-4}$) 
  and the orbits of non-spinning pulsars with otherwise same orbital parameters. 
The spinning and non-spinning pulsars follow different equations of motion 
  (i.e. choice of $\lambda$ and $C_Q$ in in Eq.~\ref{eq:mpd-momentum}),
  but the orbits they follow are indistinguishable before they fall towards
  the BH. 
During the scattering, the spin-orbit and spin-spin couplings allow the scattering angle to vary by a small amount, 
  which we denote $\Delta \phi_{\rm spin}$ and $\Delta \theta_{\rm spin}$. 
We compare this difference in scattering angle $\Delta \phi_{\rm spin}$ with the analytical formula for equatorial motion calculated by \citet{Bini2017b}, 
  and show that our results are consistent at linear order when the spin of the MSP is perpendicular to the orbital plane, 
  validating our numerical calculations. 
We would like to emphasise 
  that the results presented in this work do not include 
  explicit treatment of self-force. 
As analysed in the Sec.~\ref{sec:SelfForce}, 
  the leading order contribution of the self-force leads to negligible modification of the scattering angle
  for the systems of our interest (i.e. $v_{\infty} \ll c$).
However, this contribution does not necessarily converge for higher order self-force effects.
For example, the contribution of $\chi_{3 {\rm PM}}$ is larger than $\chi_{2 {\rm PM}}$ by
  a few order of magnitude.

The deflection angle is about $10^{-6}-10^{-3} (M/ 10^3 \solarmass)$ 
  for a typically close ($r_{\min} \le 200\, M$) fly-by, depending 
  on the orientation of the spin axis. 
Such a deflection angle can lead to a spatial difference of $30\, {\rm km}$ when the MSP reaches 
  $r/M = 20 - 20000$, much smaller than the typical influence radius of a BH with mass between 
  $10^3-4 \times 10^6 \solarmass$. 
This criterion allows us to ignore the gravitational interaction of the MSP with surrounding stars during the 
  scattering process. 
When the deviation between geodesic and non-geodesic orbits reaches $30\,{\rm km}$, 
  the predicted pulse-arrival-time using different \ac{eom} can reach $\sim 0.1\, {\rm ms}$, 
  which is about $1/10$ of the period, and definitely within the sensitivity limits of the current pulsar timing of \acp{msp}. 
Further, we show that an observation of few hours of the MSP (if we observed the MSP's signal both before and after the scattering) 
  can readily tell this differences in \ac{eom} and the effects of spin for most $r_{\min} \le 200 \, M$,
  for a intermediate-mass BH with $M=10^3 \solarmass$. 
BHs with larger masses require longer observational time ($\propto M$), but at the same time,
  has larger scattering cross sections ($\propto M^2$), which, in a naive approximation, can lead to much larger event rate.

While $1\,{\rm ms}$ 
  is about the shortest period among the 
  currently observed \acp{msp},  
  pulsars with smaller spin can also be used to measure the effects of its spin and therefore test the \ac{mpd} equations 
  (and perhaps also to set tighter constraints 
  to the equations of state of neutron stars).  
We have shown that pulsars with period $10\,{\rm ms} $ require hours or days of observations 
  (again, if we observed the pulsar's signal both before and after the scattering), 
  to measure a difference in pulse-arrival-time of $0.1\,{\rm ms}$. 
For a $10{\rm -hour}$ observation,
  the effective scattering cross section of a $10\,{\rm ms}$ pulsar (to probe a $30\,{\rm km}$ difference in position)
  is about half of that of an MSP with the same $v_\infty \ll c$ and $M=10^3 \solarmass$.

Although no such events have been found so far, 
  it does not imply that the event rate is negligible. 
The null detection may be an artefact, 
  because existing techniques based on Fourier transformation and phase folding will likely fail for such semi-periodic radio signals.  
Detection of such a system requires a different set of templates that include GR effects in all aspects. 
We show in this work that, including the effects of the MSP's spin via the \ac{mpd} equations 
  is essential for constructing the templates for the search for such a radio signal.

\section*{Acknowledgements}

KJL is supported 
  by a PhD Scholarship from the 
  Vinson and Cissy Chu Foundation 
  and by a UCL MAPS Dean's Prize. 
KW thanks the hospitality of the CUHK Department of Physics 
 during his visits.  
This research has made use of NASA's Astrophysics Data System. 

\section*{Data availability}

The data underlying this article will be shared on reasonable request to the corresponding author.

\bibliographystyle{mnras}
\bibliography{reference.bib}  


\appendix
\section{Linear-in-spin analytic formula}
\label{ap:AnalyticalFormula}

We assume that the spinning MSP (i.e. $\lambda=1$ and $C_Q \neq 0$) and a non-spinning pulsar (i.e. $\lambda=0$ and $C_Q=0$) follow
  the same trajectory before the scattering. 
That is to say, the 4-velocity of a spinning and non-spinning MSP coincide 
  at large $r \ge 10^4 \, M$ before the scattering. 
The 4-velocity of a spinning MSP is given in Eq.~(22) of \citet{Bini2017b} (we corrected a typo in the equations):
\begin{equation}
    \begin{aligned} \label{eq:Bini4velocity}
\frac{\Delta}{M^{2}} \frac{d t}{d \tau}=& \hat{E}\left(\frac{r^{2}}{M^{2}}+\hat{a}^{2}\right)+\frac{M}{r}[(2 \hat{a}+3 s) a \hat{E}-(2 \hat{a}+s) \hat{J}] \\
&-\frac{M^{3}}{r^{3}} \hat{a}^{2} s(\hat{J}-\hat{a} \hat{E}) \ , \\
\left(\frac{d r}{d \tau}\right)^{2}=& \hat{E}^{2}-1+\frac{2 M}{r}+\frac{M^{2}}{r^{2}}\left[\hat{a}^{2}\left(\hat{E}^{2}-1\right)-\hat{J}(\hat{J}-2 \hat{E} s)\right] \\
&+\frac{2 M^{3}}{r^{3}}(\hat{J}-\hat{a} \hat{E})(\hat{J}-\hat{a} \hat{E}-3 \hat{E} s) \\
&+\frac{2 M^{5}}{r^{5}} \hat{a} s(\hat{J}-\hat{a} \hat{E})^{2} \ , \\
\frac{\Delta}{M} \frac{d \phi}{d \tau}=& \hat{J}-\hat{E} s-\frac{2 M}{r}(\hat{J}-\hat{a} \hat{E}-\hat{E} s)-\frac{M^{3}}{r^{3}} \hat{a} s(\hat{J}-\hat{a} \hat{E}) \ , 
    \end{aligned}
\end{equation}
where $\hat{a}$ is the dimensionless spin of the Kerr BH $\hat{a} = a/M$, 
  $\hat{J}$ is the dimensionless angular momentum $\hat{J} = J/M$, 
  and $\hat{E}$ is the energy per unit mass for the pulsar. 
This 4-velocity equals the 4-velocity of the non-spinning pulsar 
  whose constants of motion are:
\begin{equation}
    \begin{aligned} \label{eq:geodesicsELz}
E_0 = & \frac{1}{\sqrt{1- v_\infty^2} } \ , \\
J_0 = & \frac{\sqrt{\Delta_{\min} r_{\min} \left( (E_0^2 - 1) r_{\min}+2 M\right)}- 2 a E_0 M}{r_{\min}-2 M} \ , \\
    \end{aligned}
\end{equation}  
where $\Delta_{\min}=a^2-2 M r_{\min}+r_{\min}{}^2$. 
Therefore, we have 
\begin{equation}
    \begin{aligned}
\hat{E} = & \frac{E_0 \left(r_0^3-a M^2 s\right)+J_0 M^2 s}{r_0^3-M^3 s^2} \ , \\
\approx & E_0+ \frac{M^2}{r_0^3} s  \left(J_0-a E_0\right) + \mathcal{O} \left(s^2\right) \ , \\
\hat{J} = &  \frac{E_0 M s \left(r_0^3-a^2 M\right)+J_0 \left(a M^2 s+r_0^3\right)}{M r_0^3-M^4 s^2} \ , \\
\approx & \frac{J_0}{M}+s \left(-\frac{a^2 E_0 M}{r_0^3}+\frac{a J_0 M}{r_0^3}+E_0\right)+ \mathcal{O}\left(s^2\right)  \ . 
    \end{aligned}
\end{equation}  
Further, our $(e', p')$ satisfy 
\begin{equation}
    \begin{aligned}
\frac{d r}{d \tau} \bigg|_{r = M p' /(1+e')}  = \frac{d r}{d \tau} \bigg|_{r = M p' /(1-e')} = 0 \ , \\
    \end{aligned}
\end{equation}
and can be found by either solving these two equations with Eq.~\ref{eq:Bini4velocity},
  or equivalently solving Eq.~(26-31) in \citet{Bini2017b}.
It is a set of non-linear equations, 
  and the solution does not have a closed form. 
Therefore, we calculate $(e', p')$ numerically throughout the paper.

Similarly, the geodetic $e$ and $p$ are defined such that 
\begin{equation}
    \begin{aligned}
\frac{d r}{d \tau} \bigg|_{r = M p /(1+e)}  = \frac{d r}{d \tau} \bigg|_{r = M p /(1-e)} = 0 \ , \\
    \end{aligned}
\end{equation}
with $\hat{E} \to E_0$, $\hat{J} \to J_0/M$, $s\to 0$ in Eq.~\ref{eq:Bini4velocity}.

\section{Accuracy of the deflection angle}
\label{sec:Accuracy}

\begin{figure}
  \centering 
  \includegraphics[width=1\columnwidth]{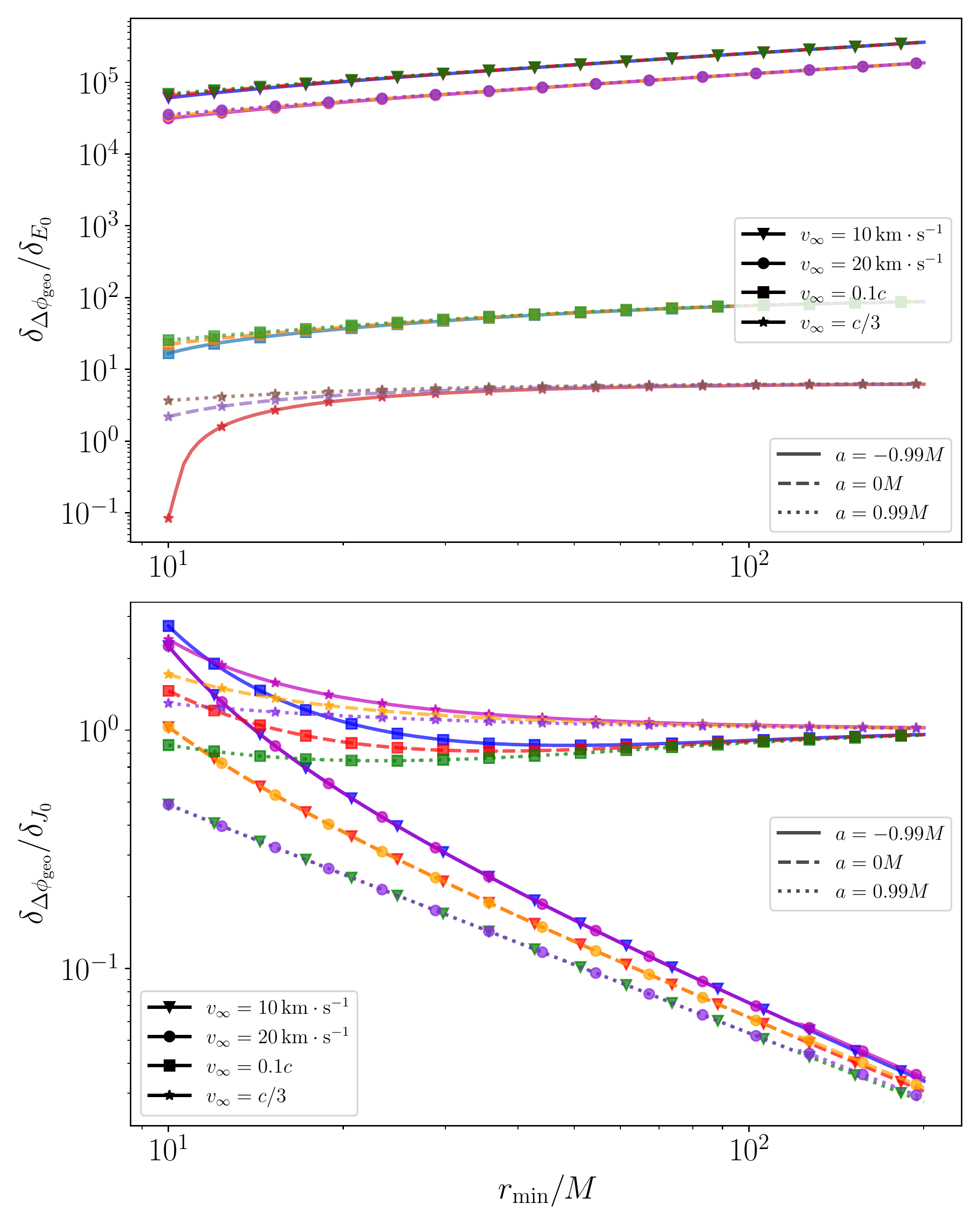}
  \vspace{-0.7cm} 
  \caption{The variation of the deflection angle $\Delta \phi_{\rm geo}$
  when the initial energy $E_0$ or angular momentum $J_0$ is perturbed by a small amount.}
  \label{fig:Perturbation}
\end{figure}

For hyperbolic orbits with $v_{\infty} \ll c$,
  the orbit transits from hyperbolic to parabolic when $v_{\infty}$ is reduced to zero. 
Therefore, the deflection angle is sensitive to the value of $v_{\infty}$
  and hence the value of $E_0$.
To illustrate this, the variation of geodetic deflection angle $\Delta {\phi_{\rm geo}}$
  is calculated when $E_0$ or $J_0$ is perturbed by a small value.
We use $\delta_{E_0}$, $\delta_{J_0}$ and $\delta_{\Delta {\phi_{\rm geo}}}$  to denote the ratio variation of $E_0$, $J_0$ and $\Delta {\phi_{\rm geo}}$, respectively. 
As shown in Fig.~\ref{fig:Perturbation},
  perturbing $J_0$ by $\delta_{J_0}$ would only lead to $\delta _{\Delta {\phi_{\rm geo}}}$
  of comparable order. 
However, when $E_0$ is perturbed, 
  the ratio variation of $\Delta \phi_{\rm geo}$ is, in general, larger than $\delta_{E_0}$.
For $v_{\infty} \ll c$, $\delta_{\Delta {\phi_{\rm geo}}}$ is larger than $\delta_{E_0}$ 
  by a few order of magnitude, and is approximately inversely proportional to $v_{\infty}$. 
For the main system ($v_{\infty} = 10 \, \kms$) that is considered in this paper, 
  a small variation of $\delta_{E_0} = 5 \times 10^{-15}$ 
  can lead to a variation of about $ 10^{-9} \, {\rm rad}$ on $\Delta \phi_{\rm geo}$. 
For the same reason, the integration of the EOMs should be performed carefully. 
A seemingly small variation in the energy can lead to unphysical results in the deflection angle. 

\bsp	
\label{lastpage}
\end{document}